\documentclass[aps,prl,twocolumn,
superscriptaddress
]{revtex4-1} 
\usepackage{amsmath}
\usepackage{amsfonts, amssymb,amsxtra,textcomp}
\usepackage[]{graphicx}
\usepackage{color}

\usepackage{hyperref}
\usepackage[figure,table]{hypcap}               
\hypersetup{
pdftitle = {},
pdfsubject = {},
pdfauthor = {},
pdfkeywords = {},
pdfcreator = {},
pdfproducer = {LaTeX with hyperref},
colorlinks = true,
linkcolor = blue, 
anchorcolor = blue,
citecolor = blue, 
filecolor = red, 
menucolor = red, 
urlcolor  = blue, 
breaklinks = true,
pdfstartview = FitV,
pdfhighlight = /I,
pdfpagelayout = OneColumn,
hypertexnames=true 
}

\makeatletter
\@ifundefined{textcolor}{}
{%
 \definecolor{BLACK}{gray}{0}
 \definecolor{WHITE}{gray}{1}
 \definecolor{RED}{rgb}{1,0,0}
 \definecolor{GREEN}{rgb}{0,1,0}
 \definecolor{BLUE}{rgb}{0,0,1}
 \definecolor{CYAN}{cmyk}{1,0,0,0}
 \definecolor{MAGENTA}{cmyk}{0,1,0,0}
 \definecolor{YELLOW}{cmyk}{0,0,1,0}
}

 \@ifundefined{textcolor}{}{%
  \definecolor{BLACK}{gray}{0}
  \definecolor{WHITE}{gray}{1}
  \definecolor{RED}{rgb}{1,0,0}
  \definecolor{GREEN}{rgb}{0,1,0}
  \definecolor{BLUE}{rgb}{0,0,1}
  \definecolor{CYAN}{cmyk}{1,0,0,0}
  \definecolor{MAGENTA}{cmyk}{0,1,0,0}
  \definecolor{YELLOW}{cmyk}{0,0,1,0}
 }

 \@ifundefined{textcolor}{}{
  \definecolor{BLACK}{gray}{0}
  \definecolor{WHITE}{gray}{1}
  \definecolor{RED}{rgb}{1,0,0}
  \definecolor{GREEN}{rgb}{0,1,0}
  \definecolor{BLUE}{rgb}{0,0,1}
  \definecolor{CYAN}{cmyk}{1,0,0,0}
  \definecolor{MAGENTA}{cmyk}{0,1,0,0}
  \definecolor{YELLOW}{cmyk}{0,0,1,0}
 }
 \@ifundefined{definecolor}{
 }{}
 \@ifundefined{definecolor}{color}{}

\newcommand{\be}{\begin{equation}}
\newcommand{\ee}{\end{equation}}
\newcommand{\bea}{\begin{eqnarray}}
\newcommand{\eea}{\end{eqnarray}}
\newcommand{\bse}{\begin{subequations}}
\newcommand{\ese}{\end{subequations}}

\newcommand{\bs}[1]{\boldsymbol{#1}}

\usepackage{color}
\definecolor{d_red}{cmyk}{0.00, 0.81, 1.00, 0.27}
\definecolor{d_orange}{cmyk}{0.00, 0.33, 1.00, 0.00}
\definecolor{d_blue}{cmyk}{0.78, 0.47, 0.00, 0.20}
\definecolor{d_lgreen}{cmyk}{0.07, 0.00, 0.79, 0.29}
\definecolor{d_green}{cmyk}{0.66, 0.00, 0.71, 0.56}
\definecolor{d_blue}{cmyk}{0.78, 0.47, 0.00, 0.20}
\definecolor{d_dblue}{cmyk}{0.91, 0.79, 0.00, 0.22}
\definecolor{d_pink}{cmyk}{0.0, 0.79, 0.37, 0.29}
\definecolor{d_purple}{cmyk}{0.16, 0.54, 0.00, 0.70}
\definecolor{d_paleblue}{cmyk}{0.669, 0.338, 0.00, 0.373}
\definecolor{d_dpaleblue}{cmyk}{0.441, 0.290, 0.00, 0.580}
\definecolor{d_brown}{cmyk}{0.0, 0.490, 0.930, 0.350}
\definecolor{d_turquoise}{cmyk}{0.630, 0.04, 0.0, 0.440}
\definecolor{KIT-green}{RGB}{0, 150,130}
\definecolor{KIT-blue}{RGB}{70,100,170}

\newcommand{\av}[1]{\langle #1 \rangle}

\newcommand{\bfbb}{{\boldsymbol{B}}}

\newcommand{\bfee}{{\boldsymbol{E}}}
\newcommand{\bfff}{{\boldsymbol{F}}}
\newcommand{\bfgg}{{\boldsymbol{G}}}

\newcommand{\bfzz}{{\boldsymbol{Z}}}

\newcommand{\bfj}{{\boldsymbol{j}}}
\newcommand{\bfk}{{\boldsymbol{k}}}

\newcommand{\bfq}{{\boldsymbol{q}}}

\newcommand{\bfv}{{\boldsymbol{v}}}

\newcommand{\bfx}{{\boldsymbol{x}}}

\newcommand{\im}{\text{Im}}

\def\bmx{\begin{pmatrix}}
\def\emx{\end{pmatrix}}
\usepackage{subfiles}
    
\begin{document}

\title{Out-of-Bounds Hydrodynamics in Anisotropic Dirac Fluids}

\author{Julia M. Link }
\affiliation{Institute for Theory of Condensed Matter, Karlsruhe Institute of
Technology (KIT), 76131 Karlsruhe, Germany}

\author{Boris N. Narozhny }
\affiliation{Institute for Theory of Condensed Matter, Karlsruhe Institute of
Technology (KIT), 76131 Karlsruhe, Germany}
\affiliation{National Research Nuclear University MEPhI
(Moscow Engineering Physics Institute), 115409 Moscow, Russia}

\author{Egor I. Kiselev }
\affiliation{Institute for Theory of Condensed Matter, Karlsruhe Institute of
Technology (KIT), 76131 Karlsruhe, Germany}

\author{J\"org Schmalian}
\affiliation{Institute for Theory of Condensed Matter, Karlsruhe Institute of
Technology (KIT), 76131 Karlsruhe, Germany}
\affiliation{Institute for Solid State Physics, Karlsruhe Institute of Technology (KIT), 76131 Karlsruhe, Germany}

\date{\today }
\begin{abstract}
We study hydrodynamic transport in two-dimensional, interacting
electronic systems with merging Dirac points at charge neutrality.
The dispersion along one crystallographic direction is Dirac-like,
while it is Newtonian-like in the orthogonal direction. As a result,
the electrical conductivity is metallic in one and insulating in the
other direction. The shear viscosity tensor contains  six independent 
components, which can be probed by measuring an anisotropic thermal 
flow. One of the viscosity components vanishes at zero temperature 
leading to a generalization of the previously conjectured lower bound 
for the shear viscosity to entropy density ratio.
\end{abstract}
\maketitle

\emph{Introduction}.--Hydrodynamic flow in quantum many-body systems is essential in systems
as diverse as superfluid helium
\cite{The-Superfluid-Phases-of-Helium3}, (Al,Ga)As heterostructures
\cite{PhysRevB.51.13389}, cold atomic ``gases''
\cite{PhysRevA.87.023629,ENSS2011770,Fluid-Dynamics-and-Viscosity-in-Strongly-Correlated-Fluids},
and the quark-gluon plasma
\cite{SHURYAK2004273,Fluid-Dynamics-and-Viscosity-in-Strongly-Correlated-Fluids}.
Recently, it has become possible to study in greater detail the
hydrodynamic flow of electrons \cite{Gurzhi,GurzhiUFN} via transport
measurements in graphene, yielding a breakdown of the Wiedemann-Franz
law \cite{Science351.1058}, superballistic transport
\cite{Falkovich,Geim17}, negative local resistance
\cite{Science351.1055,Levitov2016}, and giant magnetodrag
\cite{PhysRevLett.111.166601} (for a recent review see
Refs.~\onlinecite{ANDP:ANDP201700043,LucasFong2018}). Other key examples are
ultrapure PdCoO$_{2}$ \cite{Science351.1061} and Weyl semimetals
\cite{2017arXiv170605925G}. The appeal of these experiments is that
they allow for an investigation of the universal collision-dominated
dynamics of the pure electron fluid, largely independent of its
couplings to the lattice and impurities: the hydrodynamic flow is
expected when electron-electron scattering dominates over impurity and
electron-phonon scattering processes \cite{ANDP:ANDP201700043}.

Hydrodynamics is also one of the most successful applications of the
duality between strongly coupled gauge theories and gravity theory
\cite{1998AdTMP...2..231M}, leading to the lower bound
\cite{PhysRevLett.94.111601} for the ratio of the shear viscosity and
entropy density
\begin{equation}
\eta/s\geqslant\hbar/(4\pi k_{{\rm B}})
\:.
\label{eq:lowerbound}
\end{equation}
While originally derived as an equality for a specific
strongly coupled field theory, the bound was conjectured to apply to
all single-component nonrelativistic fluids
\cite{PhysRevLett.94.111601}. Thus, to identify a scenario where
Eq.~\eqref{eq:lowerbound} is explicitly violated is of fundamental
importance. It is also of practical relevance as a small viscosity
implies a strong tendency towards turbulent flow \cite{dau10}. 
Equation~\eqref{eq:lowerbound} can already be rationalized
using Boltzmann transport theory: Let ${s\approx
  k_{\rm{B}}\lambda^{-d}}$ be the entropy density (in $d$ dimensions)
in terms of the thermal de Broglie wavelength $\lambda$ and
${\eta\approx\epsilon\tau\lambda^{-d}}$ the shear viscosity with
energy density $\epsilon$ and scattering time $\tau$. A quasiparticle
description of transport suggests that ${\epsilon\tau\geqslant\hbar}$,
leading to the above bound (up to numerical coefficients of order
unity). A more formal reasoning can be made using scaling arguments.
We rescale distances according to
${\bs{x}\rightarrow\bs{x}'=\bs{x}/b}$ with the scaling factor
$b$. Momentum conservation and hyperscaling for critical systems imply
that both the viscosity and entropy density change according to
${\eta\left(T\right)=b^{-d}\eta\left(b^{z}T\right)}$ and
${s\left(T\right)=b^{-d}s\left(b^{z}T\right)}$, with the dynamic
scaling exponent $z$. Thus, the entropy density and shear viscosity
have the same scaling dimension. If the system approaches a fixed
point, the ratio $\eta/s$ should approach a universal value in
units of $\hbar/k_{B}$. This is analogous to the electrical
conductivity in $d=2$ that approaches a universal value in units of
$e^{2}/\hbar$, a result that follows from
$\sigma\left(T\right)=b^{2-d}\sigma\left(b^{z}T\right)$.

The bound \eqref{eq:lowerbound} appears to be violated in gravity
theories dual to an anisotropic version of a super-Yang-Mills plasma
\cite{PhysRevLett.108.021601,Jain2015,doi:10.1142/S0217984911027315}
with applications to cold gases \cite{2016arXiv160704799S}. It is of 
great interest to identify a solid-state system where such a violation
might occur.

In this Letter, we analyze the hydrodynamic behavior in an anisotropic
Dirac system, where two Dirac cones merge in momentum space
\cite{PhysRevLett.116.076803}.  Such a model is relevant to the
organic conductor $\alpha$-(BEDT-TTF)$_{2}$I$_{3}$ under pressure
\cite{PhysRevB.84.075450} and the heterostructure of the $5/3$
TiO$_{2}$/VO$_{2}$ supercell
\cite{PhysRevLett.102.166803,PhysRevLett.103.016402}. Similar behavior
is expected in the surface modes of topological crystalline insulators
with unpinned surface Dirac cones \cite{Fang2015} and quadratic double
Weyl fermions \cite{Huang2016}. In the collision-dominated regime at
charge neutrality, we predict extremely anisotropic electrical
transport exhibiting either insulating or metallic behavior depending
on the orientation of the applied electric field relative to the
crystallographic axes. Similarly, the electronic shear viscosity
strongly depends on the flow direction, exhibiting fundamentally
different temperature behavior. As a result, at low temperatures the
viscosity to entropy density ratio may diverge, stay constant, or
vanish, depending on the spatial direction. In the latter
case, Eq.~\eqref{eq:lowerbound} is violated, an effect
with experimentally measurable consequences through viscous
thermal Hagen-Poiseuille flow. We explain the anisotropic transport
in terms of the emergence of multiple length scales. In addition we 
propose a generalization of the viscosity bound to two-dimensional 
anisotropic systems:
\begin{equation}
\frac{\eta_{\alpha\beta\alpha\beta}}{s}\geqslant\frac{1}{4\pi}
\frac{\hbar}{k_{B}}\frac{\sigma_{\alpha}}{\sigma_{\beta}},
\qquad
\frac{\eta_{\alpha\beta\beta\alpha}}{s}\geqslant\frac{1}{4\pi}\frac{\hbar}{k_{B}},
\label{eq:newbound}
\end{equation}
which includes the electrical conductivity anisotropy as an additional
observable. The numerical coefficient $1/4\pi$ is consistent with
Ref.~\onlinecite{Jain2015}.

\emph{The model.}--Anisotropic Dirac systems are described by the model Hamiltonian,
${H=H_{0}+H_{C}}$, where the single-particle part is 
\begin{equation}
H_{0}=\!\int\!d^{2}r\,\psi_{r}^{\dagger}\left(-\frac{1}{2m}\nabla_{x}^{2}\sigma_{x}-iv\nabla_{y}\sigma_{y}\right)\psi_{r},\label{eq:Hamiltonian}
\end{equation}
and $H_{{\rm C}}$ is the electron-electron Coulomb interaction with
two-body potential
${V\left(r\!-\!r'\right)=e^{2}/\left|r\!-\!r'\right|}$. Here, $v$ is
the velocity along the $y$ direction and the Pauli matrices
$\sigma_{x,y}$ describe the pseudospin space. The dispersion in the
$x$ direction is characterized by the mass $m$ or, alternatively, by
the momentum scale ${k_{0}=mv}$. The anisotropy in
Eq.~\eqref{eq:Hamiltonian} is enforced by the underlying
lattice. Specifically, the direction of the parabolic dispersion (the
$x$ direction) is along the axis of the two merging Dirac points; see
Fig.~\ref{fig:fig1}. In the organic conductor
$\alpha$-(BEDT-TTF)$_{2}$I$_{3}$ \cite{PhysRevB.84.075450}, the two
Dirac cones merge together upon applying uniaxial pressure. According
to Ref.~\onlinecite{PhysRevB.84.075450}, an anisotropic Dirac cone is
expected to form at ${P=40}$~kbar; see Fig.~\ref{fig:fig1}. From
Ref.~\onlinecite{Hirata2016}, it follows that the dimensionless
strength of the Coulomb interaction is of order unity, which is
important to reach a sufficiently wide temperature regime where
electron-electron scattering dominates.
\begin{figure}
\centering
\includegraphics[width=0.85\columnwidth]{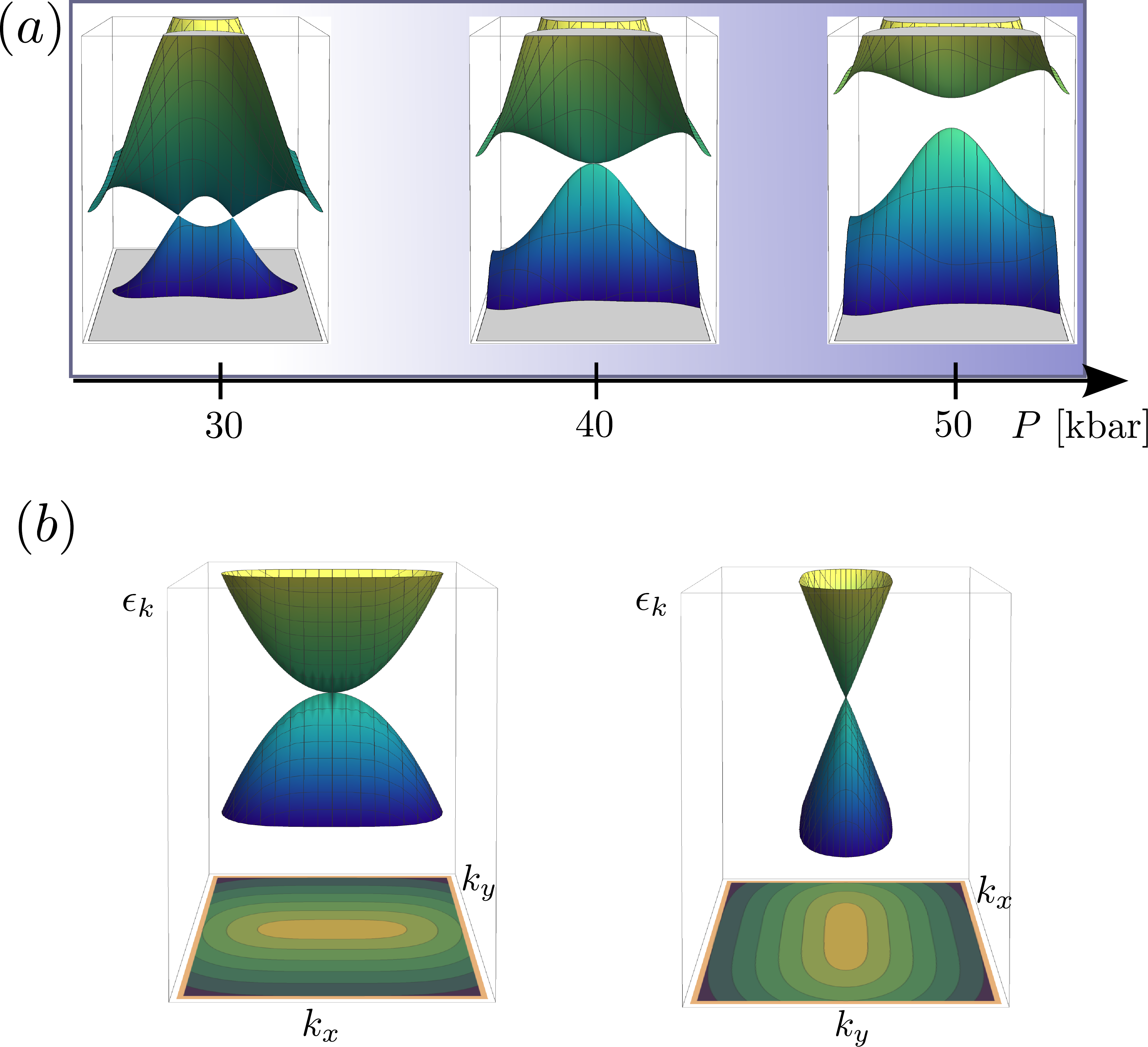} \caption{
  $(a)$ Merging Dirac cones of the organic conductor
  $\alpha$-(BEDT-TTF)$_{2}$I$_{3}$ under the application of the
  uniaxial pressure \cite{PhysRevB.84.075450}. At ${P_{a}=40}$~kbar,
  the two Dirac cones merge resulting in an anisotropic
  dispersion. $(b)$ Energy dispersion of the
  Hamiltonian~\eqref{eq:Hamiltonian}. }
\label{fig:fig1} 
\end{figure}
The renormalization group behavior of this model was recently
investigated in Ref.~\cite{PhysRevLett.116.076803} within a large-$N$
expansion \cite{PhysRevB.77.195413,PhysRevB.75.235423} ($N$ is the
number of fermion flavors; ${N\!=\!2}$ for the organic charge transfer
salts and ${N\!=\!8}$ for the oxide interfaces). While an analysis is
possible for arbitrary values of the coupling constant
${\alpha=e^{2}/\left(\hbar{v}\right)}$, we focus here on the strong
coupling behavior. In this regime, the flow equations are
\cite{PhysRevLett.116.076803}
\begin{equation}
\frac{d\alpha}{d\log b}=-\frac{0.362}{N}\alpha, \qquad
\frac{dk_{0}}{d\log b}=\frac{0.2374}{N}k_{0}.
\end{equation}
This gives rise to two characteristic length scales 
\begin{equation}
\lambda_{x}\propto T^{-\phi/z},
\qquad
\lambda_{y}\propto T^{-1/z},
\end{equation}
with the dynamic scaling exponent, $z=1-0.362/N$.
The additional
crossover exponent ${\phi=(1\!-\!0.2374/N)/2}$ 
is a measure of the anisotropy. $z<1$ reflects an increase of the
velocity at low energies and ${\phi<1/2}$ implies that interactions
make the anisotropy even stronger if compared to the bare spectrum of
Eq.\eqref{eq:Hamiltonian}. The fact that ${\phi\neq1}$ is the most
crucial ingredient of our subsequent discussion. The violation of the
lower bound only requires $\phi \neq 1$ as one viscosity component
vanishes faster than $s$ for $T \to 0$ even if the large-$N$ approach
turns out to be quantitatively inaccurate.

\emph{Scaling.}--Now we consider the constitutive relations in
anisotropic systems. The electrical conductivity is a rank-two tensor
defined in the standard way,
${j_{\alpha}=\sigma_{\alpha\beta}E_{\beta}}$ (with
${\alpha,\beta\in\left\{x,y\right\}}$). The viscosity is a rank-four
tensor connecting the dissipative part of the stress tensor
$\tau_{\alpha\beta}$ and the flow velocity gradient,
${\tau_{\alpha\beta}\!=\!\sum_{\gamma\delta}\eta_{\alpha\beta\gamma\delta}\partial_{\gamma}u_{\delta}}$.
The number of independent conductivity and viscosity coefficients can
be found from symmetry arguments \cite{SWB-074898019}. In a
rotationally invariant system, both the conductivity and shear
viscosity are each characterized by a single independent coefficient:
$\sigma_{\alpha\beta}=\sigma\delta_{\alpha\beta}$ and
$\eta_{\alpha\beta\gamma\delta}=\eta\left(\delta_{\alpha\gamma}\delta_{\beta\delta}+\delta_{\alpha\delta}\delta_{\beta\gamma}-\delta_{\alpha\beta}\delta_{\gamma\delta}\right)$
\cite{SWB-074898019}. In this Letter, we focus on incompressible
fluids and hence do not consider the bulk viscosity. In contrast, the
Hamiltonian~\eqref{eq:Hamiltonian} is not rotationally invariant and
is characterized by two conductivity elements $\sigma_{xx}$ and
$\sigma_{yy}$ and six independent viscosity coefficients with
$\eta_{\alpha \beta \gamma \delta}= \eta_{\gamma \delta \alpha
  \beta}$, such that
\begin{eqnarray}
\!\!\!\!\!\!\!\!\begin{pmatrix}\tau_{xx}\\
\tau_{xy}\\
\tau_{yx}\\
\tau_{yy}
\end{pmatrix}\!=\!\begin{pmatrix}\eta_{xxxx} & 0 & 0 & \eta_{xxyy}\\
0 & \eta_{xyyx} & \eta_{xyxy} & 0\\
0 & \eta_{yxyx} & \eta_{yxxy} & 0\\
\eta_{yyxx} & 0 & 0 & \eta_{yyyy}
\end{pmatrix}\!\begin{pmatrix}\partial_{x}u_{x}\\
\partial_{y}u_{x}\\
\partial_{x}u_{y}\\
\partial_{y}u_{y}
\end{pmatrix}\!.\label{eq:energy-stress-tensor}
\end{eqnarray}
Below we show that the off-diagonal momentum relaxation along the
$y$ direction (with linear dispersion) due to a flow with velocity
along the $x$ direction (of parabolic dispersion) described by
$\eta_{xyxy}$ is clearly different from the opposite case,
$\eta_{yxyx}$.

The scaling behavior of the conductivity and the viscosity follows
from the Kubo formalism \cite{PhysRevB.86.245309,Principi2016}. If
one takes charge conservation into account, the scaling dimension of
the conductivity is a purely ``geometric'' effect that involves the
length scales $\lambda_{\alpha}$. If $b$ is a scaling parameter for
length scales along the $y$ direction, it follows that
\cite{Sheehy2007,PhysRevLett.116.076803}
\begin{eqnarray}
\sigma_{xx}\left(T\right) & = & b^{\phi-1}\sigma_{xx}\left(b^{z}T\right),\nonumber \\
\sigma_{yy}\left(T\right) & = & b^{1-\phi}\sigma_{yy}\left(b^{z}T\right).
\end{eqnarray}
Fixing the coefficient $b$ via $b^{z}T=const$ reveals that $b^{\phi-1}$ is given by the
ratio of the two length scales,
$b^{\phi-1}\rightarrow$$\lambda_{x}/\lambda_{y}$.  It immediately
follows that $\sigma_{yy}\propto T^{(\phi-1)/z}$ diverges as
$T\rightarrow0$, while $\sigma_{xx}\propto T^{(1-\phi)/z}$
vanishes. The system is insulating along the direction with parabolic
dispersion and metallic in the other direction. Below, we confirm this
behavior within an explicit kinetic theory and see that this behavior
is pronounced below the characteristic temperature $k_{{\rm B}}
T_{0}\equiv mv^{2} \approx 1.5$~eV in the organic salts.

Similar behavior emerges for the entropy density and the components of
the viscosity tensor. For the entropy density, it follows from the
usual scaling behavior of anisotropic systems \cite{Hornreich1975},
$s\left(T\right)=b^{-\left(1+\phi\right)}s\left(b^{z}T\right)$,
yielding
${s\left(T\right)\propto{k}_{B}/\left(\lambda_{x}\lambda_{y}\right)}$.
To determine the behavior of the viscosity tensor we use again the
Kubo formalism (the details are summarized in the Supplemental
Material \cite{supplemental}). The result is that most tensor elements
$\eta_{\alpha\beta\gamma\delta}$ have the same scaling dimension as
$s\left(T\right)$. However, there are two crucial exceptions:
\begin{eqnarray}
\eta_{xyxy}\left(T\right) & = & b^{-\left(3-\phi\right)}\eta_{xyxy}\left(b^{z}T\right),\nonumber \\
\eta_{yxyx}\left(T\right) & = & b^{-\left(3\phi-1\right)}\eta_{yxyx}\left(b^{z}T\right).
\label{eq:two_special_visc_coeff}
\end{eqnarray}
It immediately follows that $\eta_{xyxy}/s\propto
T^{2\left(1-\phi\right)/z}$ and $\eta_{yxyx}/s\propto
T^{-2\left(1-\phi\right)/z}$. Unless the system is isotropic and
${\phi=1}$, one component vanishes and the other diverges.  Thus,
regardless of the numerical coefficient of $\eta_{xyxy}$, it will
violate the bound Eq.~\eqref{eq:lowerbound} at sufficiently low
temperatures since $\phi<1$. Below we obtain this behavior from
Boltzmann theory as well. The scaling analysis reveals that the charge
transport and momentum transport are closely related to each
other. This allows us to construct combinations of physical
observables that have scaling dimension zero. These combinations are
listed in Eq.~\eqref{eq:newbound} and give rise to the generalized
lower bound for the viscosity tensor.

\emph{Hydrodynamics.}--The scaling behavior can be obtained from the kinetic
equation 
\begin{equation}
\frac{\partial f_{\mu\bfk}}{\partial t}+\bfv_{\mu\bfk}\cdot\frac{\partial f_{\mu\bfk}}{\partial\bfx}
+e\bs{E}\cdot\frac{\partial f_{\mu\bfk}}{\partial\bfk}=I_{\mu}^{ee},\label{eq:Boltzmann}
\end{equation}
where $f_{\mu\bfk}$ is the distribution function for a quasiparticle
from the band $\mu$ and with the quasimomentum $\bfk$, and
$I_{\mu}^{ee}$ is the collision integral due to the Coulomb
interaction. The latter we treat in perturbation theory in $1/N$ (for
details see Refs.~\onlinecite{supplemental,RPA}).

Following the standard derivation of the hydrodynamic theory in the
limit of an incompressible fluid \cite{dau10}, we integrate the
kinetic equation (\ref{eq:Boltzmann}) and obtain generalizations of
the Navier-Stokes equation at the charge neutrality point. Flow along
the direction of the parabolic dispersion is described by a
Navier-Stokes equation similar to that of a Galilean invariant system,
\begin{equation}
m_{*}n \left(\partial_{t}u_{x}+\!u_{i}\partial_{i}u_{x}\right)+\partial_{x}P=\mathcal{F}_{s,x}+m_{*}n \partial_{i}\delta j_{I}^{i},
\end{equation}
where $m_{*}\approx1.37m$, $n$ is the total quasiparticle density,
$\delta\bfj_{I}$ is the dissipative correction to the total
quasiparticle current, $P$ is the hydrodynamic pressure, and
$\mathcal{F}_{s,\beta}=\partial_{\alpha}\tau_{\alpha\beta}=\partial_{\alpha}\eta_{\alpha\beta\gamma\delta}\partial_{\gamma}u_{\delta}$
is the Stokes force. Flow in the $y$ direction obeys the equation
similar to that in graphene \cite{PhysRevLett.103.025301,us2}
\begin{eqnarray}
 &  & Ts\left(\partial_{t}u_{y}+u_{i}\partial_{i}u_{y}\right)+\partial_{y}P+u_{y}\partial_{t}P\\
 &  & \qquad=\mathcal{F}_{s,y}-u_{y}E_{i}\delta j_{i}+u_{y}\partial_{x_{i}}\delta j_{\epsilon,i}
 \:.
 \nonumber 
\end{eqnarray}
The dissipative terms include the corrections to the electric current,
$\delta\bfj$, and the energy current, $\delta\bfj_{\epsilon}$.
\begin{figure}
\includegraphics[scale=0.5]{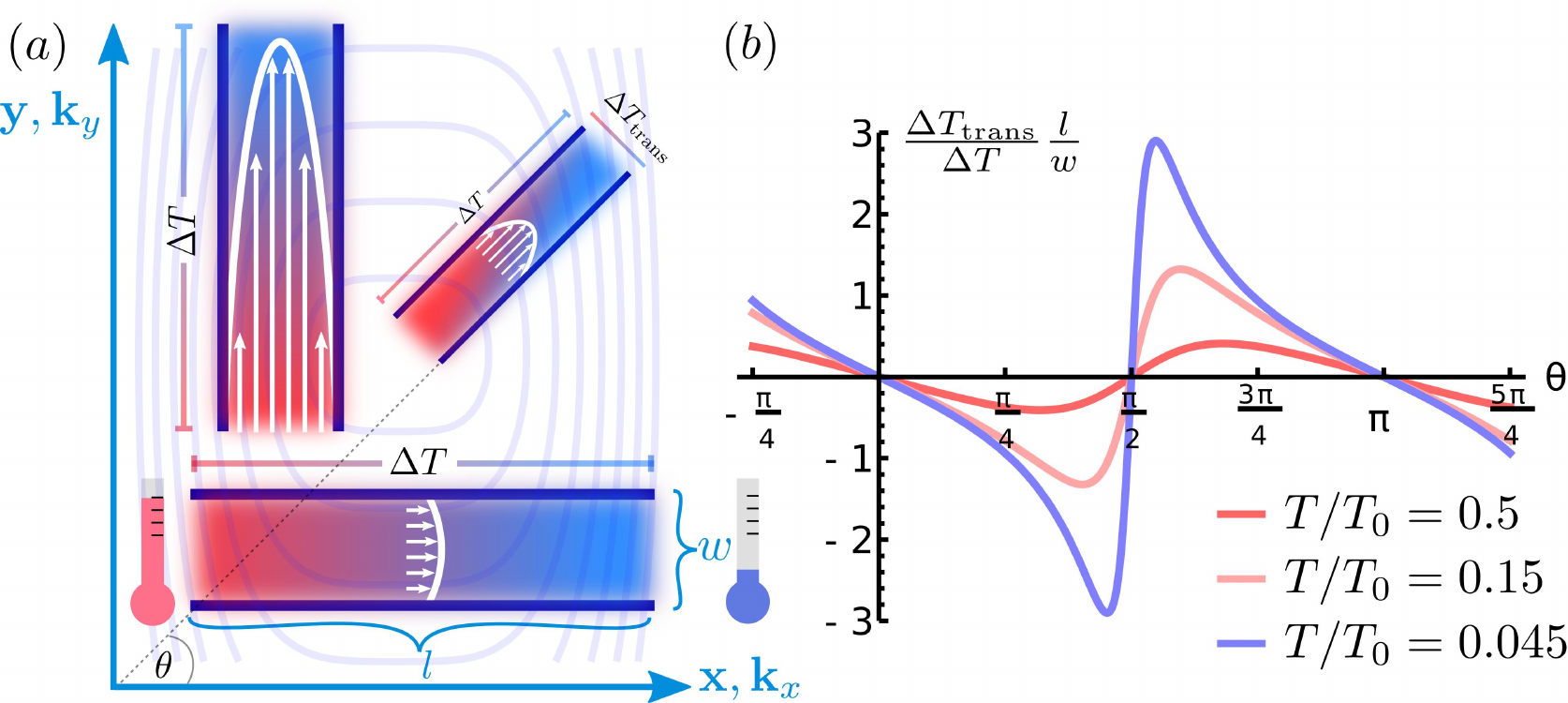}
\caption{(a) Hagen-Poiseuille flow profile due to a
  temperature gradient.  The bound-violating small ratio
  $\eta_{xyxy}/s$ leads to a parabolic flow profile with large
  curvature and yields a large thermal conductivity, while the large
  ratio $\eta_{yxyx}/s$ yields almost Ohmic flow. (b) Transverse
  temperature gradient as function of the angle $\theta$ between flow
  direction and $x$ axes (see left panel).}
\label{fig:fig2} 
\end{figure}

The simplest nontrivial solution of these equations can be obtained in
the linearized stationary regime in the absence of the electric
field. Using Eq.~\eqref{eq:energy-stress-tensor}, we find
\begin{equation}
\partial_{\alpha}P=\tilde{\eta}_{\alpha\alpha\alpha\alpha}\partial_{\alpha}^{2}u_{\alpha}
+2 \tilde{\eta}_{\alpha\alpha\overline{\alpha}\overline{\alpha}}\partial_{\alpha}\partial_{\overline{\alpha}}u_{\overline{\alpha}}
+\tilde{\eta}_{\overline{\alpha}\alpha\overline{\alpha}\alpha}\partial_{\overline{\alpha}}^{2}u_{\alpha}
\label{eq: Hagen Pois}
\end{equation}
with $\tilde{\eta}_{\alpha \beta \gamma \delta}=\frac{1}{2}(\eta_{\alpha \beta \gamma \delta}+\eta_{\gamma \beta \alpha \delta})$, 
and ${\overline{x}\!=\!y}$ and vice versa. Note that $\tilde{\eta}_{xyxy}=\eta_{xyxy}$ and the same 
for $\tilde{\eta}_{yxyx}$. From the kinetic equation
it also follows that the heat current,
${\bs{j}_{\epsilon}\approx(5/3)\epsilon\bs{u}}$, is solely determined
by the flow velocity. This is similar to the particle current in
Galilean-invariant systems and reflects the fact that the thermal
conductivity of a Dirac system at neutrality is infinite in the limit
of infinite size \cite{PhysRevB.78.115406,Foster2009}. Using
${\partial_{\alpha}P\!=\!-s\partial_{\alpha}T}$ at neutrality, we
can solve the above equations for a finite geometry, find the velocity
profile $\bs{u}$, and determine the thermal current from
$\bs{j}_{\epsilon}$. Note that this is how the entropy
density enters our theory. 

Consider now a flow in a system of width $w$. In this geometry, there
is no net flow in the lateral $x$ direction. The solution of
Eq.~\eqref{eq: Hagen Pois} with the no-slip boundary conditions,
${u_{y}\left(x=\pm{w}/2\right)=0}$, yields the standard
Hagen-Poiseuille profile
\begin{equation}
u_{y}\left(x\right)=\frac{s}{2\eta_{xyxy}}
\left(\frac{w^2}{4}-x^{2}\right)\partial_{y}T.
\end{equation}
Integration over the cross section yields the total thermal current
${I_{\epsilon}=w\kappa_{yy}\partial_yT}$ with the thermal
conductivity $\kappa_{yy}=(5\epsilon w^{2}/24)(s/\eta_{xyxy})$.
Similar analysis of the flow along the $x$ direction yields
$\kappa_{xx}=(5\epsilon w^{2}/24)(s/\eta_{yxyx})$. In the case of
no-stress boundary conditions it is better to analyze conical flow.

The above results demonstrate that the thermal Hagen-Poiseuille 
flow is determined precisely by those viscosity tensor
elements~\eqref{eq:two_special_visc_coeff} that violate the ordinary
scaling behavior. Thus, the ratios $\eta_{xyxy}/s$ and $\eta_{yxyx}/s$
matter and are in fact the easiest to observe. The other tensor
components emerge only when the flow direction is not aligned with one
of the crystalline axes. In this case a transverse temperature
gradient builds up, see Fig.~\ref{fig:fig2}.
\begin{figure}
\includegraphics[width=\columnwidth]{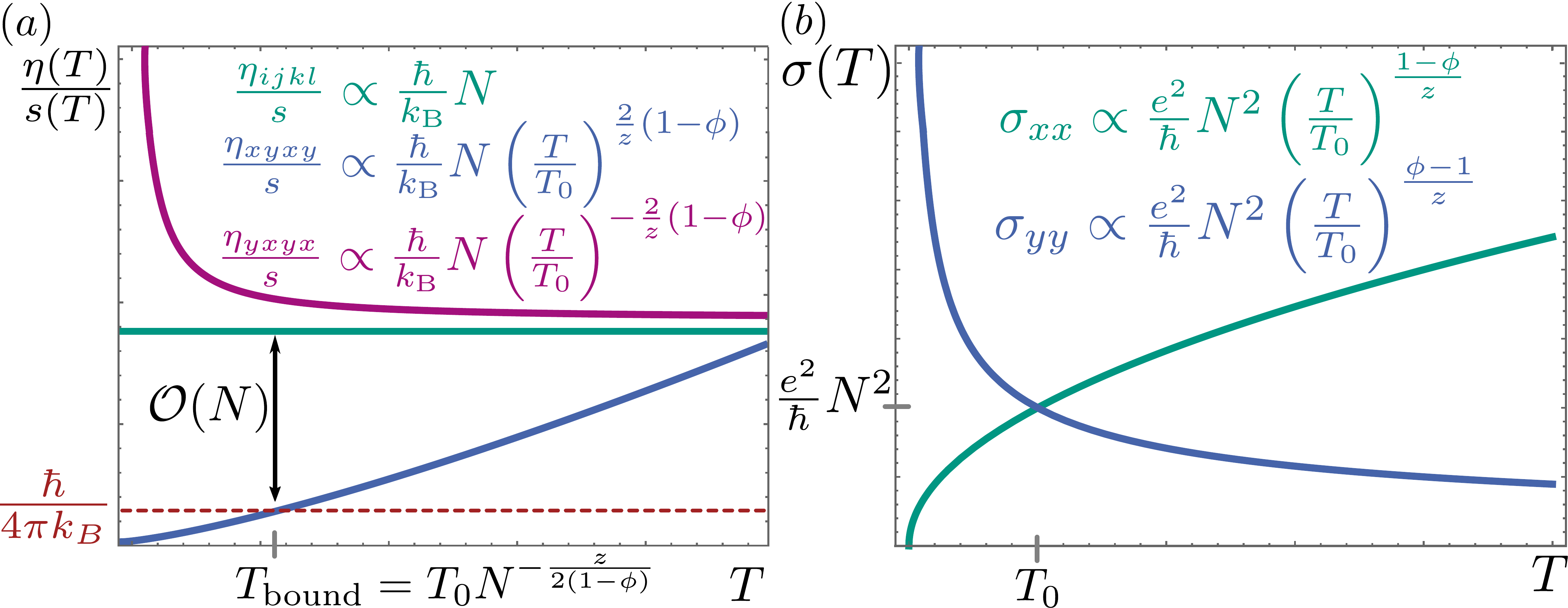} \caption{(a) Ratio $\eta_{\alpha\beta\gamma\delta}/s$ as
a function of temperature. The viscosity coefficient $\eta_{xyxy}$
violates the lower bound (shown by the dashed line). (b)
$\sigma_{\alpha\alpha}$ as function of temperature with insulating
and metallic conductivity along the direction with parabolic and linear
dispersion, respectively.}
\label{fig:fig3} 
\end{figure}

\emph{Kinetic theory.} Finally, we use the microscopic quantum kinetic
equation approach to find the conductivities and viscosities. The
former can be found in the standard way \cite{dau10}: applying a weak
electric field, $\bs{E}$, we drive the system out of equilibrium where
the distribution function $f_{\mu\bs{k}}$ acquires a nonequilibrium
correction proportional to the field:
$\delta{f}_{\mu\bs{k}}\!=\!f_{\mu\bfk}^{(0)}(1\!-\!f_{\mu\bfk}^{(0)})h_{\mu\bfk}/T$,
${h_{\mu\bfk}\!=\!\mu\bfv_{\mu\bfk}\bfee{g}_{\mu\bfk}^{\bfee}}$. Solving
the kinetic equation for the functions $g_{\mu\bfk}^{\bfee}$, we find
$\sigma_{xx,yy}(T)\propto{N}^{2}(e^{2}/\hbar)(T/T_{0})^{\pm(1-\phi)/z}$
in agreement with the scaling results.

To determine the viscosity, we have to find the stress tensor within
linear response to the external shear force. In terms of the
nonequilibrium distribution function, the stress tensor is given by
${\tau_{\alpha\beta}\!=\!\sum_{\mu}\int_{k}\!v_{\mu\bfk}^{\alpha}k_{\beta}\delta{f}_{\mu\bfk}}$.
The correction $\delta{f}_{\mu\bfk}$ is proportional to the velocity
gradients with
$h_{\mu\bfk}\!=\!\sum_{\alpha\beta}\mu(v_{\mu\bfk}^{\alpha}k_{\beta}\!-\!\delta_{\alpha\beta}\epsilon_{\mu\bfk}/2)\partial_{\alpha}{u}_{\beta}g_{\mu\bfk}^{\beta}$.
Now we expand the functions $g_{\mu\bfk}^{\beta}$ in the basis of the
eigenfunctions of the linearized kinetic equation
\cite{us2,PhysRevB.91.035414},
${g_{\mu\bfk}^{\beta}\!=\!\sum_{n}\psi_{n}^{\beta}\phi_{\mu\bfk}^{(n)}}$. The
dominant contribution comes from the two modes describing the energy
and the energy band index. This allows us to represent the kinetic
equation \eqref{eq:Boltzmann} in the matrix form,
${\mathcal{M}_{u_{\beta,\alpha}}^{ee}\boldsymbol{\psi}^{\beta}\!=\!\bfgg_{u_{\beta,\alpha}}}$,
where the matrix $\mathcal{M}_{u_{\beta,\alpha}}^{ee}$ corresponds to
the collision integral due to Coulomb interaction. The exact form of
$\mathcal{M}_{u_{\beta,\alpha}}^{ee}$ and $\bfgg_{u_{\beta,\alpha}}$
can be found in Ref.~\onlinecite{supplemental}. Solving the matrix
equation, we find $g_{\mu\bfk}^{\beta}$ and hence the viscosity
coefficients,
$\eta_{\alpha\beta\gamma\delta}=\sum_{\mu}\int_{k}\mu{v}_{\mu\bfk}^{\alpha}k_{\beta}(v_{\mu\bfk}^{\gamma}k_{\delta}-\delta_{\gamma\delta}\epsilon_{\mu\bfk}/2)g_{\mu\bfk}^{\beta}f_{\mu\bfk}^{(0)}(1-\!f_{\mu\bfk}^{(0)})/T$.
At charge neutrality, the resulting viscosities are given by
${\eta_{\alpha\beta\gamma\delta}=N^{2}\mathcal{C}_{\alpha\beta\gamma\delta}(k_0^2/\hbar)(T/T_{0})^{\phi_{\alpha\beta\gamma\delta}}}$,
where $\mathcal{C}_{\alpha\beta\gamma\delta}$ are numerical
coefficients of order unity. The exponents
$\phi_{\alpha\beta\gamma\delta}$ coincide with the results of the
above scaling analysis.

The linear-response solution of the quantum kinetic equation yields
the entropy density in the scaling form,
$s\!=\!N\mathcal{C}_{s}k_{{\rm B}}(k_0^2/\hbar^2)
(T/T_0)^{(1+\phi)/z}$. Because of the linear dispersion in the
$y$ direction, the velocity component $v_{y}$ at the scale $T$ is
larger than $v_{x}$. The viscosity coefficient $\eta_{yxyx}$ describes
the flow of the momentum $k_{x}$ with the larger velocity component
$v_{y}$ leading to the diverging ratio ${\eta_{yxyx}/s}$. In contrast,
the viscosity coefficient $\eta_{xyxy}$ corresponds to the flow of the
momentum $k_{y}$ with the much slower velocity $v_{x}$, leading to the
violation of the bound. The ratios of the viscosity coefficients to
the entropy density are shown in Fig.~\ref{fig:fig3}.

\emph{Summary.} In this letter, we have shown that anisotropic Dirac
systems are fascinating new materials with unparalleled transport
properties in the hydrodynamic regime. The same material exhibits both
insulating and metallic behavior depending on the direction of the
applied electrical field. Furthermore, the shear viscosity is represented 
by a fourth rank tensor with six independent components with the ratio 
$\eta_{\alpha\beta\gamma\delta}/s$ that may either vanish or diverge, see Fig.~\ref{fig:fig3}.
In the former case, the universal bound (\ref{eq:lowerbound}) appears 
to be violated.  We demonstrated that these viscosity tensor elements can be measured via
viscous thermal flow, where the more electrically
conducting direction is also the direction of larger thermal
conductivity. The thermal flow in the direction with the linear
spectrum is expected to be highly susceptible to turbulence and should
lead to large transverse temperature variations. [ see Fig.2(b)]. 
Similar behavior will also occur in other anisotropic systems,
such as critical bosonic systems at a Lifshitz point
\cite{Hornreich1975}. The ratio $\eta/s$ (\ref{eq:lowerbound}) was introduced to have a
measure for the strength of interaction in a quantum fluid. Our
analysis shows that the violation of the bound for anisotropic systems
is not necessarily an indicator for extreme interactions, but reflects
the fact that $\eta/s$ is no longer an appropriate measure of the
interaction strength. We have suggested a generalization 
of the lower bound that takes into account the anisotropy. The generalized bound~\eqref{eq:newbound} offers a
better quantification of fluid interactions. Nevertheless, the smallness 
of the viscosity to entropy density ratio $\eta/s \ll
\hbar/(4 \pi k_B)$ for anisotropic systems remains a strong indicator
for a tendency towards turbulent flow. 

We gratefully acknowledge illuminating discussions with A. V. Chubukov,
I. V. Gornyi, K. Kanoda, J. Klier, A. D. Mirlin, M. M{\"u}ller, D.
E. Sheehy, and T. Neupert. J.M.L. thanks the Carl-Zeiss-Stiftung for
financial support. B.N.N. acknowledges support by the EU Network Grant
FP7-PEOPLE-2013-IRSES 612624 ``InterNoM'' and the MEPhI Academic
Excellence Project (Contract No. 02.a03.21.0005). 
The work of J.S. was performed in part at the Aspen Center for Physics, 
which is supported by NSF Grant No. PHY-1607611.


%
%
%

    \newpage

\title{Supplemental Material to Out-of-bounds Hydrodynamics in anisotropic Dirac systems}
\author{Julia M. Link }
\affiliation{Institute for Theory of Condensed Matter, Karlsruhe Institute of
Technology (KIT), 76131 Karlsruhe, Germany}

\author{Boris N. Narozhny }
\affiliation{Institute for Theory of Condensed Matter, Karlsruhe Institute of
Technology (KIT), 76131 Karlsruhe, Germany}
\affiliation{National Research Nuclear University MEPhI
(Moscow Engineering Physics Institute), 115409 Moscow, Russia}

\author{Egor I. Kiselev }
\affiliation{Institute for Theory of Condensed Matter, Karlsruhe Institute of
Technology (KIT), 76131 Karlsruhe, Germany}

\author{J\"org Schmalian}
\affiliation{Institute for Theory of Condensed Matter, Karlsruhe Institute of
Technology (KIT), 76131 Karlsruhe, Germany}
\affiliation{Institute for Solid State Physics, Karlsruhe Institute of Technology (KIT), 76131 Karlsruhe, Germany}

\date{\today}
\begin{abstract}
In this Supplemental Material details to the calculation of the results presented in ``Out-of-bounds Hydrodynamics in anisotropic Dirac systems'' are given.
The collision integral due to Coulomb interaction is shown as well as the explicit expressions needed for the determination of the viscosity
and conductivity tensor. There is also a section describing the derivation of the Navier-Stokes equation in more detail and a further section in which the scaling
behavior of the conductivity and the viscosity tensor is derived using the Kubo formalism.
\end{abstract}

\maketitle
\section{The collision integral}
The anisotropic Dirac systems (ADSs) are described by
the Hamiltonian ${H=H_{0}+H_{C}}$, where the single-particle part is
\begin{equation}
  H_{0}= \!\int\! d^2r \, \psi_r^{\dagger}
  \left(-\frac{1}{2m}\nabla_{x}^{2}\sigma_{x}-iv\nabla_{y}\sigma_{y}\right)\psi_r,
\label{eq:Hamiltonian}
\end{equation}
and $H_C$ is the Coulomb interaction. Here, $v$ is the velocity along
the $y$-direction and the Pauli matrices $\sigma_{x,y}$ describe the
pseudo-spin space corresponding to the conductance and valence bands.
The $x$-direction we will characterize by the momentum scale
${k_{0}=2mv}$. The eigenenergies of the system are
\begin{equation}
\label{ea}
\epsilon_{\lambda\bs{k}} = \lambda v_a \sqrt{\frac{k_x^4}{k_0^2}+k_y^2},
\end{equation}
leading to the following expression for the quasiparticle velocities
\begin{equation}
\label{va}
\bs{v}_{\lambda \bs{k}}=\frac{v_a^2}{\epsilon_{\lambda\bs{k}}} \left(2\frac{k_x^2}{k_0^2} k_x, k_y\right).
\end{equation}

In order to determine transport properties of the system such as the
conductivity $\sigma$ and the shear viscosity $\eta$ we use the
kinetic (Boltzmann) equation \cite{SWB-074898019}:
\begin{equation}
  \frac{\partial f_{\lambda\bfk}}{\partial t}+
  \bfv_{\lambda\bfk} \frac{\partial f_{\lambda\bfk}}{\partial \bfx} +
  \bfff \frac{\partial f_{\lambda\bfk}}{\partial \bfk}
 = - \frac{\delta f_{\lambda\bfk}}{\tau} +I_{\lambda}^{ee},
  \label{eq:Boltzmann}
\end{equation}
where $f_{\lambda\bfk}$ is the distribution function for a
quasiparticle from the band $\lambda$ and with the quasimomentum
$\bfk$,
${\delta{f}_{\lambda\bfk}=f_{\lambda\bfk}-\langle{f}_{\lambda\bfk}\rangle_\varphi}$
(where the angular average is performed over the directions of
$\bfk$), ${\tau}$ is the relaxation time due to impurity scattering,
${\bfff=e(\bfee\!+\!\bfv\!\times\!\bfbb/c)}$ is the Lorentz force, and
$I_{\lambda}^{ee}$ is the collision integral due to the Coulomb
interaction. The latter is given by
\begin{eqnarray}
	I_{\lambda}^{ee}
	&=&
	e^2 \sum_{q,\mu} \im D(\bfq, \epsilon_{\lambda\bfk} - \epsilon_{\mu\bfk-\bfq})
        N_{\lambda \mu}(\bfk, \bfk-\bfq)
        \\
        &&
        \nonumber\\
        &\times&
        \big[n( \epsilon_{\lambda\bfk}-\epsilon_{\mu\bfk-\bfq})
          \left( f_{\lambda\bfk}-f_{\mu\bfk-\bfq}\right)
        + f_{\lambda\bfk} (1-f_{\mu\bfk-\bfq}) \big],
        \nonumber
\end{eqnarray}
with
\[
N_{\lambda \mu}(\bfk_1,\bfk_2)\!=\!\frac{1}{2}
\left[1\!+\!\frac{\lambda \mu }{\epsilon_{\lambda\bfk_1} \epsilon_{\mu\bfk_2}}
  \left( \frac{k_{1,x}^2 k_{2,x}^2}{k_0^2}+ k_{1,y} k_{2,y} \right) \right].
\]
The function ${\im D(\bfq,\omega)\!=\!\im \left(q \!-\! 2 \pi e^2
\Pi\left(\omega,q \right) \right)^{-1}}$ is the spectral function of
the Coulomb propagator. In the strong coupling limit, it
simplifies to $\im \Pi(\omega,q)^{-1}$. The polarization operator
$\Pi\left(\omega,q \right)$ is given by
\begin{equation}
	\Pi(\omega,q)
	=
	-\alpha
	\left[\frac{d_x \left(v_a/k_0 \right)^{1/2} q_x^2}{\Delta(\omega,\bfq)^{1/4}} + \frac{d_y \left( v_a/k_0 \right)^{-1/2} q_y^2}{\Delta(\omega,\bfq)^{3/4}} \right]
	\:,
\end{equation}
where $\Delta(\omega,\bfq) = \omega^2 + c \left( v_a/k_0 \right)^{2} q_x^4+v_a^2 q_y^2$ and $d_x,d_y,$ and $c$ are constants (see Ref.~\onlinecite{PhysRevLett.116.076803}).

Next we linearize the collision integral introducing the standard
nonequilibrium correction to the local equilibrium distribution
function $f_{\lambda\bfk}^{(0)}$
\[
\delta{f}_{\lambda\bfk}\!=\!f^{(0)}_{\lambda\bfk}(1\!-\!f^{(0)}_{\lambda\bfk})h_{\lambda\bfk}/T.
\]
Thus, in the strong coupling limit the linearized collision integral
is
\begin{widetext}
\begin{eqnarray}
	I_{\lambda}^{ee}
	&=&
	\frac{1}{T}\sum_q \im \Pi\left(\bfq, \epsilon_{\lambda\bfk} - \epsilon_{\lambda\bfk-\bfq}\right)^{-1}
        N_{\lambda \lambda}(\bfk, \bfk-\bfq)  
	n^{(0)}(\epsilon_{\lambda\bfk} - \epsilon_{\lambda\bfk-\bfq})
        \left[1-f_{\lambda\bfk}^{(0)} \right] f_{\lambda\bfk-\bfq}^{(0)}
        \left[ h_{\lambda\bfk}-h_{\lambda\bfk-\bfq}\right]
        \\
	 &+&
	\frac{1}{T}\sum_q \im \Pi(\bfq, \epsilon_{\lambda\bfk} + \epsilon_{\lambda\bfk-\bfq})^{-1}
        N_{\lambda -\lambda}(\bfk, \bfk-\bfq) 
	n^{(0)}(\epsilon_{\lambda\bfk} + \epsilon_{\lambda\bfk-\bfq})
        \left[1-f_{\lambda\bfk}^{(0)} \right]f_{-\lambda\bfk-\bfq}^{(0)}
        \left[ h_{\lambda\bfk}-h_{-\lambda\bfk-\bfq}\right],
        \nonumber
\end{eqnarray}
where $n^{(0)}(\epsilon)$ is the bosonic equilibrium distribution
function.  The first term of the collision integral describes
intraband scattering processes while the second term arises due to
interband scattering.

\end{widetext}


%
\section{The Navier-Stokes equation}
Starting from the Boltzmann equation Eq.~\eqref{eq:Boltzmann}, we can derive the 
continuity equation for the particle density, the energy density and the momentum density.
But before we start, let us define the different densities of the two-band system.
The number of carriers in the two bands are defined as
%

\begin{subequations}
\label{den}
\begin{equation}
\label{np}
n_+=N\!\int\!\!\frac{d^2k}{(2\pi)^2} f_{+,\bfk},
\end{equation} 
and
\begin{equation}
\label{nm}
n_-=N\!\int\!\!\frac{d^2k}{(2\pi)^2} \left(1-f_{-,\bfk}\right),
\end{equation}
with the total ``charge'' (or ``carrier'') density being
\begin{equation}
\label{n}
n = n_+ - n_-.
\end{equation}
If the densities of the conduction and valence bands are conserved
independently (e.g., in graphene), then one can also define the
``imbalance'' or the total quasiparticle density
\begin{equation}
\label{ni}
n_I = n_+ + n_-.
\end{equation}
\end{subequations}
The energy density is
\begin{equation}
  n_\epsilon =N\sum\limits_\lambda
  \!\int\!\!\frac{d^2k}{(2\pi)^2} \epsilon_{\lambda\bfk} f_{\lambda\bs{k}} - n_{\epsilon0},
\end{equation}
where we subtract the energy of the filled valence band
\[
n_{\epsilon0 }= N\!\int\!\!\frac{d^2k}{(2\pi)^2} \epsilon_{-,\bfk}.
\]
The momentum density has the form
\begin{equation}
\bs{n}_{\bs{k}} = N\!\sum_\lambda\!\int\!\frac{d^2k}{(2\pi)^2} \bs{k} f_{\lambda\bs{k}}.
\end{equation}
The corresponding continuity equations are given by
\begin{subequations}
\label{ces}
\begin{equation}
\label{cen1}
\partial_t n + \bs{\nabla}_{\bs{r}}\!\cdot\!\bs{j} = 0,
\end{equation}
\begin{equation}
\label{cene1}
\partial_t n_E + \bs{\nabla}_{\bs{r}}\!\cdot\!\bs{j}_\epsilon = e \bs{E}\cdot\bs{j},
\end{equation}
\begin{equation}
\label{cek1}
\partial_t n^\alpha_{\bs{k}} + \nabla^\beta_{\bs{r}} \Pi_{\beta\alpha}
- e n E^\alpha - \frac{e}{c} \left[\bs{j}\!\times\!\bs{B}\right]^\alpha =
- \frac{n^\alpha_{\bs{k}}}{\tau_{\rm dis}}.
\end{equation}
\end{subequations}

In order to calculate the above densities and currents, we 
consider the local equilibrium distribution function
\begin{equation}
\label{le}
f^{(0)}_{\lambda\bs{k}} (\bs{r}) =
\left\{
1+\exp\left[\frac{\epsilon_{\lambda\bs{k}}-\mu_\lambda(\bs{r}) - 
\bs{u}(\bs{r})\!\cdot\!\bs{k}}{T(\bs{r})}\right]
\right\}^{-1},
\end{equation}
where ${\mu_\lambda(\bs{r})}$ is the local chemical potential and
$\bs{u}(\bs{r})$ is the hydrodynamic (or ``drift'') velocity.
Furthermore, we expand $f^{(0)}_{\lambda\bs{k}}$ in the power series
in small $\bs{u}$ (as compared to either $v$ or the velocity of the
parabolic direction at the energy scale $T$),
\begin{equation}
\label{lex}
f^{(0)}_{\lambda\bs{k}}\approx f^{(F)}_{\lambda\bs{k}} 
- \bs{u}\!\cdot\!\bs{k} \frac{\partial f^{(F)}_{\lambda\bs{k}}}{\partial\epsilon_{\lambda\bs{k}}}
+\frac{1}{2} (\bs{u}\!\cdot\!\bs{k})^2
\frac{\partial^2 f^{(F)}_{\lambda\bs{k}}}{\partial\epsilon^2_{\lambda\bs{k}}}
+\dots
\end{equation}
Using this expansion, we evaluate the
electrical current (up to the electron charge $e$),
 \begin{equation}
  \bs{j} = n \bs{u} +\delta \bs{j},
 \end{equation}
 and the energy current,
 \begin{equation}
  \bs{j}_\epsilon = \frac{5}{3} n_\epsilon \bs{u} +\delta \bs{j}_\epsilon,
 \end{equation}
where $\delta \bs{j}$ and $\delta \bs{j}_\epsilon$ represent the dissipative
corrections to the (energy)-current defined by the out-of-equilibrium
distribution function.

The energy density at the charge neutrality point has the temperature
dependence
\begin{equation}
 n_\epsilon(\mu=0)= \frac{2 \sqrt{k_0}}{\pi^2 v^{3/2}} K(-1) N Z_{3/2} T^{5/2},
\end{equation}
with
$Z_{3/2}= \frac{3}{4} \left(1-\frac{\sqrt{2}}{4} \right) \sqrt{\pi} \zeta(5/2)$.
The momentum density of our ADSs is 
\begin{equation}
  n_{\bfk}^x = m_* n_I u_x,
  \qquad
 n_{\bfk}^y = \frac{5}{3 v^2} n_\epsilon u_y,
\end{equation}
where 
${m_*=3[E(-1)-K(-1)]/K(-1)} m\approx1.37~m$ with $E$ and
$K$ being the complete elliptic integrals.

The total quasiparticle density at charge neutrality is
 \begin{equation}
  n_I=n_++n_- \stackrel{\mu \rightarrow 0}{\longrightarrow} \frac{2 \sqrt{k_0}}{\pi^2 v^{3/2}} K(-1) N Z_{1/2} T^{3/2}\:,
 \end{equation}
with
$Z_{1/2}=\frac{1}{2} \left(1-\frac{1}{\sqrt{2}} \sqrt{\pi} \zeta(3/2) \right)$.

The momentum flux (or stress tensor) is defined as 
\begin{equation}
\label{pab}
\Pi_{\alpha\beta} = N\!\sum_\lambda\!\int\!\!\frac{d^2k}{(2\pi)^2} 
k^\beta v^\alpha_{\lambda\bs{k}} f_{\lambda\bs{k}},
\end{equation}

Under the assumption of local equilibrium, the stress tensor has the
form:
\begin{eqnarray}
&&
\Pi_{xx} = \frac{2}{3} n_E + \frac{3}{2} m_* n_{I} u_x^2
+ \frac{5}{6 v_a^2} n_{E} u_y^2 + \tau_{xx},
\nonumber\\
&&
\nonumber\\
&&
\Pi_{yy} = \frac{2}{3} n_E + \frac{1}{2} m_* n_{I} u_x^2
+ \frac{5}{2 v_a^2} n_{E} u_y^2 + \tau_{yy},
\nonumber\\
&&
\nonumber\\
&&
\Pi_{xy} = m_* n_{I} u_xu_y + \tau_{xy},
\nonumber\\
&&
\nonumber\\
&&
\Pi_{yx} = \frac{5}{3v_a^2} n_{E} u_xu_y + \tau_{yx}.
\end{eqnarray}

For an arbitrary two-band system in the state of local equilibrium,
the pressure can be obtained by using the relation between the thermodynamical potential $\Omega$ and the pressure $P	$, i.~e. $\Omega =-P V$,
which yields
\begin{eqnarray}
\label{p}
&&
\!\!\!\!\!\! P = T N\!\int\!\frac{d^2k}{(2\pi)^2} 
\ln \left[1 + e^{\displaystyle\frac{\mu_+\!-\!\epsilon_{+,\bs{k}}\!+\!\bs{u}\!\cdot\!\bs{k}}{T}}\right]
\\
&&
\nonumber\\
&&
\qquad\quad
+ T N\!\int\!\frac{d^2k}{(2\pi)^2} 
\ln \left[1 + e^{\displaystyle\frac{-\mu_-\!-|\epsilon_{-,\bs{k}}|\!-\!\bs{u}\!\cdot\!\bs{k}}{T}}\right].
\nonumber
\end{eqnarray}
After integrating by parts, one finds for the pressure
\begin{eqnarray}
&&
P_+ \!=\! \frac{1}{2} N\!\!\int\!\!\frac{d^2k}{(2\pi)^2}
\left(\bs{k}\!\cdot\!\bs{v}_{+\bs{k}}-\bs{k}\!\cdot\!\bs{u}\right)f^{(0)}_{+\bs{k}},
\\
&&
\nonumber\\
&&
 P_- \!=\! \frac{1}{2} N\!\!\int\!\!\frac{d^2k}{(2\pi)^2}
\left(\bs{k}\!\cdot\!\bs{v}_{-\bs{k}}-\bs{k}\!\cdot\!\bs{u}\right)
\left(f^{(0)}_{-\bs{k}}-1\right),
\end{eqnarray}
which leads to the following expression connecting the pressure to the energy stress tensor:
\begin{equation}
 P = \frac{1}{2} {\rm Tr \,} \Pi_{\alpha\beta} 
- \frac{1}{2} \bs{n}_{\bs{k}}\!\cdot\!\bs{u}.
\end{equation}
For ADSs, the pressure is thus
\begin{equation}
 P=\frac{2}{3} n_\epsilon +\frac{1}{2} m_* n_I u_x^2 +\frac{5}{6 v^2} n_\epsilon u_y^2,
\end{equation}
and the enthalpy of the system is:
\begin{equation}
 w = n_\epsilon +P.
\end{equation}
Combining the expressions for the pressure and stress tensor, one can re-write the diagonal elements of the stress tensor in the form
\begin{eqnarray}
 \Pi_{xx} &=& P + m_* n_I u_x^2 +\tau_{xx}\\
 \Pi_{yy} &=& P + \frac{5}{3 v^2} n_\epsilon u_y^2 + \tau_{yy}.
\end{eqnarray}
At last we can now insert all these expressions into the continuity equation for the momentum density and obtain the 
Navier-Stokes-equation:
\begin{widetext}
\begin{eqnarray}
  m_* ~n_I \left(\partial_t u_x + u_i \partial_i u_x \right) + \partial_x P
	&=&
	\mathcal{F}_{s,x} + e E_x n + m_*~n_I~  \partial_{x_i} \delta j_i\\
  w \left( \partial_t u_y+ u_i \partial_i u_y \right) + \partial_y P +u_y \partial_t P
	&=&
	\mathcal{F}_{s,y}+ e n \left( E_y- u_y  u_i E_i\right)-u_y E_i \delta j_i +u_y \partial_{x_i} \delta j_{\epsilon,i}
	\:,
\end{eqnarray}
where $\mathcal{F}_{s,\alpha}=\partial_i \tau_{i\alpha} = \partial_j \eta_{j\alpha a b} \frac{\partial u_b}{\partial x_a}$
is the Stoke-force.
From the Navier-Stokes equation, the frequencies of the shear modes is derived which are
\begin{eqnarray}
\omega_{\pm} &=&
 i \frac{q_x^2}{2} \left( \frac{ v^2 \eta_{xyxy}}{ sT} +\frac{ \eta_{xxxx}}{m_* n_I} \right)+ i \frac{q_y^2}{2} \left( \frac{v^2 \eta_{yyyy}}{ sT} +\frac{ \eta_{yxyx}}{m_* n_I} \right)  \\
 &\pm&
 i
 \sqrt{ \frac{4 (\eta_{xxyy}+\eta_{yxxy}) (\eta_{yyxx}+\eta_{xyyx}) m_* n_I q_x^2 q_y^2 v^2 s T+\left(q_x^2 (\eta_{xxxx} s T-\eta_{xyxy} v^2 m_* n_I)+q_y^2 (\eta_{yxyx} s T-\eta_{yyyy}v^2 m_* n_I)\right)^2}{4 m_*^2 n_I^2 s^2 T^2} }
 \nonumber
\end{eqnarray}

%
%
%
%
%
\section{The viscosity tensor in linear response}

In order to determine the shear viscosity of the system, we implement
a Boltzmann equation on which no Lorentz force acts. Furthermore, we
will also neglect impurities and take only scattering due to Coulomb
interaction into account yielding the following expression
\begin{equation}
 \frac{\partial f_{\lambda\bfk}}{\partial t}+ \bfv_{\lambda\bfk} \frac{\partial f_{\lambda\bfk}}{\partial \bfx}  =
  I_{\lambda}^{ee}
  \:.
  \label{eq:Boltzmannhydro}
\end{equation}
Within linear response, the nonequilibrium correction to the
distribution function is given by
\begin{equation}
  h_{\lambda\bfk}\!=\!\sum_{\alpha\beta} \lambda
  \left(v^{\alpha}_{\lambda \bfk } k_\beta\!-\!\frac{1}{2}\delta_{\alpha\beta}\epsilon_{\lambda\bfk}\right)
  \frac{\partial{u}_\beta}{\partial{x}_\alpha}g_{\lambda\bfk}^{\beta}, 
\end{equation}
where we sum over equal indices and use the incompressibility of the
system, i.e. $\partial_i u_i=0$.
The linearized Boltzmann equation can now be written as
\begin{equation}
  \frac{\partial f_{\lambda\bfk}}{\partial t}
  + \frac{1}{T} \left( \bfv_{\lambda \bfk}^\alpha k_\beta-\frac{\delta_{\alpha \beta} \epsilon_{\lambda \bfk}}{2} \right)
  \frac{\partial u_\beta}{\partial x_\alpha}  f^{(0)}_{\lambda\bfk}\left[1-f^{(0)}_{\lambda\bfk} \right]
=I^{ee}_{\lambda}(k, h_{\lambda\bfk})
	\:.
\end{equation}

The function $g_{\lambda\bfk}^{\beta}$ can be expanded into a set of
basis function,
$g_{\lambda\bfk}^{\beta}\!=\!\sum_n\psi^\beta_n\phi^{(n)}_{\lambda\bfk}$. The
two dominant modes determining the behavior of the system describe the eigenenergy of the
system $\phi^{(0)}_{\lambda\bfk}=\lambda \epsilon_{\lambda\bfk}$ and the band index $\phi^{(1)}_{\lambda\bfk}=\lambda$.

Multiplying the above equation with a mode $\phi^{(n)}_{\lambda\bfk}$
from the left, summing over $\lambda$ and integrating over $\bfk$, we
represent the Boltzmann equation in the following matrix form
\begin{equation}
  \begin{pmatrix}
   \mathcal{M}^{ee}_{u_{x,y}} & 0\\
   0 & \mathcal{M}^{ee}_{u_{y,x}}
  \end{pmatrix}\begin{pmatrix}
		   \boldsymbol{\psi}^x\\
		    \boldsymbol{\psi}^y
		\end{pmatrix}
	=
	\begin{pmatrix}
	 \bfgg_{u_{x,y}}\\
	 \bfgg_{u_{y,x}}
	\end{pmatrix}
	\:,
\end{equation}
where the matrix $\mathcal{M}^{ee}_{u_{\beta,\alpha}}$ describing the Coulomb interaction is
%
 \begin{eqnarray}
 \mathcal{M}_{nm,~u_{\beta,\alpha}}^{ee}
 &=&
 \frac{N}{T}\!\sum_\lambda\!\int\!\!\frac{d^2k}{(2\pi)^2}
 \sum_q \im \Pi\left(\bfq, \epsilon_{\lambda\bfk} - \epsilon_{\lambda\bfk-\bfq}\right)^{-1}
 N_{\lambda \lambda}(\bfk, \bfk-\bfq)
 n^{(0)}(\epsilon_{\lambda\bfk} - \epsilon_{\lambda\bfk-\bfq})
        \left[1-f_{\lambda\bfk}^{(0)} \right] f_{\lambda\bfk-\bfq}^{(0)}
 \nonumber\\
 &&\qquad\times
 \left[ \lambda \left(v^{\alpha}_{\lambda \bfk} k_\beta\!-\!\frac{1}{2}\delta_{\alpha\beta}\epsilon_{\lambda\bfk}\right)
   \phi^{(n)}_{\lambda\bfk}\phi^{(m)}_{\lambda\bfk}-
   \lambda \left( v_{\lambda \bfk-\bfq}^\alpha (\bfk-\bfq)_\beta -\frac{1}{2} \delta_{\alpha \beta} \epsilon_{\lambda\bfk-\bfq} \right)
   \phi^{(n)}_{\lambda\bfk-\bfq}\phi^{(m)}_{\lambda\bfk}\right]
 \nonumber\\
	 &+&
 \frac{N}{T}\!\sum_\lambda\!\int\!\!\frac{d^2k}{(2\pi)^2}
 \sum_q \im \Pi(\bfq, \epsilon_{\lambda\bfk} + \epsilon_{\lambda\bfk-\bfq})^{-1}
 N_{\lambda -\lambda}(\bfk, \bfk-\bfq) 
 n^{(0)}(\epsilon_{\lambda\bfk} + \epsilon_{\lambda\bfk-\bfq})
 \left[1-f_{\lambda\bfk}^{(0)} \right]f_{-\lambda\bfk-\bfq}^{(0)}
 \nonumber\\
 &&\qquad\times
\left[\lambda \left(v^{\alpha}_{\lambda \bfk} k_\beta\!-\!\frac{1}{2}\delta_{\alpha\beta}\epsilon_{\lambda\bfk}\right)
  \phi^{(n)}_{\lambda\bfk}\phi^{(m)}_{\lambda\bfk}
  -\lambda \left( v_{\lambda \bfk-\bfq}^{\alpha} (\bfk-\bfq)_\beta-\frac{1}{2} \delta_{\alpha \beta} \epsilon_{\lambda\bfk-\bfq}\right)
   \phi^{(n)}_{-\lambda\bfk-\bfq} \phi^{(m)}_{\lambda\bfk}\right],
  \nonumber
\end{eqnarray}
and the vector $\bfgg_{u_{\beta,\alpha}}$
\begin{equation}
  G_{m,~u_{\beta,\alpha}}=\frac{N}{T} \!\sum_\lambda\!\int\!\!\frac{d^2k}{(2\pi)^2}
  \phi^{(m)}_{\lambda\bfk} f_{\lambda\bfk}^{(0)}\left[1-f_{\lambda\bfk}^{(0)} \right]
  \left(v^\alpha_{\lambda \bfk} k_\beta -\frac{1}{2} \delta_{\alpha \beta}\epsilon_{\lambda\bfk} \right).
\end{equation}
%
%
Upon introducing dimensionless variables, $\Omega = \omega/T$, $x=\sqrt{ 1/(T k_0)} ~ k_x$, and $y= k_y/T$, the temperature dependence of the different terms are found
\begin{eqnarray}
	\mathcal{M}_{nm,u_{x,y}}^{ee}
	&=&
	 T^2 \left[ \phi^{(m)}_{\lambda \bfk} \right] \left[ \phi^{(n)}_{\lambda \bfk}\right] \mathcal{C}^{ee}_{nm, u_{x,y}} \\
	\mathcal{M}_{nm,u_{y,x}}^{ee}
	&=&
	 T^3 \left[ \phi^{(m)}_{\lambda \bfk} \right] \left[ \phi^{(n)}_{\lambda \bfk} \right] \mathcal{C}^{ee}_{nm,u_{y,x}}                                                        
	\:,
\end{eqnarray}
and
\begin{eqnarray}
	G_{m,u_{x,y}}
	&=&
	T ~[\phi^{(m)}_{\lambda \bfk}] ~\mathcal{G}_{m,u_{x,y}}  \\
	G_{m,u_{y,x}}
	&=&
	T^2~ [\phi^{(m)}_{\lambda \bfk}] ~\mathcal{G}_{m,u_{y,x}}
	\:,
\end{eqnarray}
where $\mathcal{C}^{ee}_{nm,u_{\beta,\alpha}}$ and $\mathcal{G}_{m,u_{\beta,\alpha}} $ are numerical coefficients and $[\phi^{(0)}_{\lambda \bfk}]=[\epsilon_k]=T$, and $[\phi^{(1)}_{\lambda \bfk}]=[\lambda]=1$. After inversion of this matrix equation, 
we find that the temperature dependence of $g_\beta(k,\lambda)$ is proportional to the inverse temperature $T^{-1}$ and the fermionic flavor $N$. We can now determine the shear viscosity, since
%
\begin{eqnarray}
 \av{\tau_{\alpha \beta}}
 = 
 \frac{N}{T}\sum_{\lambda,\gamma,\delta}\int_k \lambda~ v_{\lambda \bfk}^{\alpha} k_\beta ~\left( v_{\lambda \bfk}^{\gamma} k_\delta-\frac{1}{2} \delta_{\gamma \delta} \epsilon_{\lambda \bfk} \right) ~ f^{(0)}_{\lambda \bfk}(1-f^{(0)}_{\lambda \bfk}) ~g^{\delta}_{\lambda \bfk} ~\frac{\partial u_\delta}{\partial x_\gamma} 
 = \sum_{\gamma \delta} \eta_{\alpha \beta \gamma \delta} \frac{\partial u_\delta}{\partial x_\gamma}.
\end{eqnarray}
The temperature dependence of the viscosity tensor is defined as
\begin{equation}
 \begin{pmatrix}
   \tau_{xy}\\
  \tau_{yx} 
 \end{pmatrix}
 =
 \begin{pmatrix}
 N^2 \left( \frac{T}{v k_0} \right)^{3/2} \frac{k^2_0}{\hbar} ~\mathcal{C}_1 & N^2 \left( \frac{T}{v k_0} \right)^{5/2} \frac{k^2_0}{\hbar} ~\mathcal{C}_2\\
 N^2 \left( \frac{T}{v k_0} \right)^{1/2} \frac{k^2_0}{\hbar} ~\mathcal{C}_3 & N^2 \left( \frac{T}{v k_0} \right)^{3/2} \frac{k^2_0}{\hbar} ~\mathcal{C}_1
 \end{pmatrix}
  \begin{pmatrix}
   \frac{\partial u_x}{\partial y}\\
   \frac{\partial u_y}{\partial x}
  \end{pmatrix}
\:.
\end{equation}
\end{widetext}
%
%
%
%
%
%
%
%
%
%
%
%
%
%
%
%
%
\newpage

\section{The electrical conductivity in linear response}
In order to determine the conductivity of the system, we assume a stationary and uniform system
\begin{equation}
 \frac{\partial f_{\lambda \bfk}}{\partial t}=\frac{\partial f_{\lambda \bfk}}{\partial \bfx}=0
\end{equation}
on which the Lorentz force
$
 \bfff= e \bfee 
$
acts with $\bfee$ being the electrical field.
We choose as ansatz for the nonequilibrium correction of the distribution function
\begin{eqnarray}
	h_{\lambda\bfk}
	&=&
	\!\lambda \bfv_{\lambda \bfk}\bfee{g}^{E}_{\lambda\bfk}
	\:,
\end{eqnarray}
where the functions $g^{E}_{\lambda \bfk}$ can be expanded again in a set of basis functions, 
$g^{E}_{\lambda \bfk}=\sum_n \psi_n^{E} \phi^{(n)}_{\lambda \bfk}$.
The linearized Boltzmann equation can now be cast into a matrix equation by multiplying it from the left with different basis functions and integrating and summing over $k$ and $\lambda$, respectively.
It has than the compact form
\begin{equation}
  \mathcal{M}^{ee}_{\bfee} 
   \boldsymbol{\psi} =
    \bfgg_{\bfee}
   \:,
\label{eq:sup:matrixeqconductivity}
\end{equation}
where the matrix $\mathcal{M}^{ee}_{\bfee}$ describes scattering processes
due to Coulomb interaction. Upon introducing dimensionless variables, $\Omega = \omega/T$, $x=\sqrt{ 1/(T k_0)} ~ k_x$, and $y= k_y/T$, 
the temperature dependence of the matrix elements is determined. 
Thus for the matrix elements describing the Coulomb interaction, we find
\begin{widetext}
\begin{eqnarray}
 \mathcal{M}_{nm}^{ee}
 &=&
 \frac{1}{T}\sum_{\lambda} \int_k \sum_q \im \Pi\left(\bfq, \epsilon_{\lambda\bfk} - \epsilon_{\lambda\bfk-\bfq}\right)^{-1} N_{\lambda \lambda}(\bfk, \bfk-\bfq)  
	 n^{(0)}( \epsilon_{\lambda \bfk} - \epsilon_{\lambda \bfk-\bfq})\left[1-f_{\lambda \bfk}^{(0)} \right] f_{\lambda \bfk-\bfq}^{(0)} \left[ \phi^{(n)}_{\lambda \bfk} \phi^{(m)}_{\lambda \bfk}-\phi^{(n)}_{\lambda \bfk -\bfq} \phi^{(m)}_{\lambda \bfk}\right] \nonumber\\
	 &+&
	\frac{1}{T}\sum_q \im \Pi\left(\bfq, \epsilon_{\lambda\bfk} + \epsilon_{\lambda\bfk-\bfq}\right)^{-1} N_{\lambda -\lambda}(\bfk, \bfk-\bfq) 
	  n^{(0)}( \epsilon_{\lambda \bfk}+ \epsilon_{\lambda \bfk-\bfq})\left[1-f^{(0)}_{\lambda \bfk} \right] f^{(0)}_{-\lambda \bfk-\bfq} \left[ \phi^{(n)}_{\lambda \bfk} \phi^{(m)}_{\lambda \bfk}-\phi^{(n)}_{-\lambda \bfk-\bfq} \phi^{(m)}_{\lambda\bfk}\right] \nonumber\\
 &=&
 T^{3/2}~ k_0^{1/2} ~[\phi^{(m)}_{\lambda \bfk}] [\phi^{(n)}_{\lambda \bfk}]~ \mathcal{C}_{nm,\bfee}^{ee}
 \:.
\end{eqnarray}
The vector $G_{m,\bfee}$ is defined as
\begin{eqnarray}
 G_{m,\bfee}
 &=&
  \frac{1}{T} ~\sum_{\lambda} \int_k  ~\lambda \phi^{(m)}_{\lambda \bfk}~ f^{(0)}_{\lambda \bfk}\left[1-f^{(0)}_{\lambda \bfk} \right]\\
  &=& T^{1/2} [\phi^{(m)}_{\lambda \bfk}] \mathcal{C}_{m,\bfee}^{f}
 \:.
\end{eqnarray}
In order to obtain the coefficients $\psi^{E}_n$, we have to invert the above matrix equation Eq.~\eqref{eq:sup:matrixeqconductivity}. These coefficients
are important since they determine the current and thus the conductivity. The current is defined as
\begin{equation}
	\bfj
	=
	\frac{e N}{T} \sum_{\lambda,n} \int_k  \lambda \bfv_{\lambda k} f^{(0)}_{\lambda \bfk} \left(1-f^{(0)}_{\lambda \bfk} \right)  v^i_{\lambda \bfk}E_i \psi_n^{E} 
	\:,
\end{equation}
or written in components as
\begin{eqnarray}
	j_\alpha 
	&=&
	\frac{e N}{T}\sum_{\lambda,n} \int_k \lambda  f^{(0)}_{\lambda \bfk} \left(1-f^{(0)}_{\lambda \bfk} \right) \left(v^{\alpha}_{\lambda \bfk}\right)^2 \psi^{E}_n E_{\alpha}  \phi^{(n)}_{\lambda \bfk}
	\:.
\end{eqnarray}
\end{widetext}
At the charge neutrality point $(\mu=0)$, the electrical conductivity can be rewritten as
\begin{eqnarray}
	\sigma_{\alpha \alpha}=\frac{j_\alpha}{E_\alpha}
	&=& 
	N \sum_\lambda \int_k  \frac{\left( v^\alpha_{\lambda \bfk}\right)^2}{T} \frac{2 e^{\epsilon_{\lambda k} /T}}{(1+e^{\epsilon_{\lambda k} /T})^2}~  \psi_1^{E}
	\:.
\end{eqnarray}
Thus the temperature dependence is
\begin{equation}
	\sigma_{{xx},{yy}}
	=
	N \left( \frac{T}{v k_0}\right)^{3/2,1/2}~ \psi^{E}_1~ \mathcal{C}^{x,y}
	\,
\end{equation}
where $\mathcal{C}^{x,y}$ are numerical coefficients. If we have no magnetic field, we find for the coefficient $\psi^{E}_1$
\begin{equation}
 \psi_1^{E}
 =
N \frac{v k_0}{T}\frac{\mathcal{C}^f_{1,\bfee}  \mathcal{C}^{ee}_{00,\bfee} }{\mathcal{C}^{ee}_{00,\bfee} ~\mathcal{C}^{ee}_{11,\bfee}-{\mathcal{C}^{ee}_{01,\bfee}}^2}
\:.
\end{equation}
 %
%
%
%
%
%
%
\section{RG and scaling behavior of physical observables}
%
%
%
%
%
%
The renormalization analysis in the large $N$-limit and the strong 
coupling regime of Ref.~\cite{PhysRevLett.116.076803} is used in this section.
When we consider an observable, we expect the scaling behavior $(b=e^l)$

\begin{equation}
		O(\bfk,T,\alpha)=Z_OO[\bfzz_\bfk(l^\ast)\bfk,Z_T(l^\ast)T,\alpha(l^\ast)]
		\:,
\end{equation}
where $Z_O$ is the scaling dimension of the operator and
\begin{eqnarray}
		\frac{d Z_{\omega}}{d l} 
		&=&
		Z_{\omega} \left( 1- \gamma_v \right)\\
		\frac{d Z_x}{d l}
		&=&
		\frac{1}{2} Z_x \left(1-\gamma_{k_0} \right)
		\:.		
\end{eqnarray}
For the coupling constant $\alpha$ and the momentum scale $k_0$, we obtain the flow equations
\begin{equation}
	\frac{d \alpha}{d l}
	=
	-\alpha \gamma_v
	\text{  ~and~  }
	\frac{d k_0}{d l}
	=
	k_0 \gamma_{k_0}
	\:,
\end{equation}
where in the strong coupling limit, it is~\cite{PhysRevLett.116.076803}
\begin{equation}
		\gamma_v
		= \frac{0.3625}{N}
		\text{  ~and~  }
		\gamma_{k_0}
		=
		\frac{0.2364}{N}
\:.
\end{equation}
This yields
\begin{eqnarray}
 \alpha(b) &=& \alpha b^{-\gamma_v}\\
 k_0(b) &=& k_0 b^{\gamma_k}
 \:,
\end{eqnarray}
i.e. the coupling constant flows from the assumed large value towards weak coupling, while the momentum scale $k_0$ becomes increasingly larger.
Furthermore, it is
\begin{eqnarray}
		Z_\omega
		&=&
		b^{1-\gamma_v}= b^z\\
		Z_x
		&=&
		b^{\frac{1}{2}\left( 1-\gamma_{k_0}\right)}= b^{\phi}
		\:.
\end{eqnarray}
%
%
\subsubsection{Compressibility}
In the case of the particle density $O=n$ follows $Z_{O=n}= b^{-1} Z_x^{-1}=b^{-(\phi+1)}$, such that the compressibility $\kappa = \frac{\partial n}{\partial \mu}$ has $Z_{\kappa}= Z_\omega/(b Z_x)=b^{z-(\phi+1)}$.
Scaling then implies
\begin{equation}
		\kappa(T) 
		=
		b^{z-(1+\phi)} \kappa\left( b^z T\right)
		\propto T^{\phi_\kappa}
		\:,
\end{equation}
with
\begin{equation}
		\phi_\kappa
		=
		\frac{1}{2} -\frac{1}{2}(\gamma_{k_0}- 3 \gamma_v) = \frac{1}{2} + \frac{0.4255}{N}
		\:.
\end{equation}
The compressibility vanishes less slowly than in the free anisotropic Dirac fermion prediction $\kappa_0 \propto T^{1/2}$.\\
\subsubsection{Conductivity}
We want to determine the scaling dimension of the electrical conductivity $\sigma_{\alpha \alpha}$ with $\alpha={x,y}$. Therefore, we consider 
the optical conductivity $\sigma_{\alpha \alpha}$ in the collisionless regime. 
The same powerlaws are expected for the $T$-dependent conductivity in the collision dominated regime. We use the Kubo-formula
\begin{equation}
	\sigma_{\alpha \alpha} 
	=
	\lim \limits_{\bfq \rightarrow 0} \frac{\omega}{q_\alpha^2} \chi\left( \bfq, \omega \right)
	\:,
\end{equation}
with the charge susceptibility $\chi$. The charge susceptibility is proportional to the compressibility in the limit of zero momentum and frequency 
\begin{equation}
	\chi\left( \bfq \rightarrow 0, \omega = 0 \right) = \kappa
	\:.
\end{equation}
Hence the charge susceptibility has the same scaling dimension as the compressibility
$
	Z_\chi=Z_\kappa
	\:.
$
This implies that we find for the conductivity the following scaling dimensions
\begin{eqnarray}
	\sigma_{xx} &=& \frac{Z_x}{b} \sigma_{xx}\left(Z_\omega T \right)
	=
	b^{\phi-1} \sigma_{xx}(b^z T) \\
	\sigma_{yy} &=& \frac{b}{Z_x} \sigma_{yy}\left(Z_\omega T \right)
	=
	b^{1-\phi} \sigma_{yy}(b^z T)
	\:.
\end{eqnarray}
We thus find the temperature dependence for the conductivity
\begin{equation}
	\sigma_{xx,yy}(T) 
	\propto
	\left( \frac{T}{T_0} \right)^{\pm \frac{1-\phi}{z}}
	\propto
	\left( \frac{T}{T_0} \right)^{\pm(\frac{1}{2}+\phi_{\sigma})}
	\;,
\end{equation}
with $\phi_{\sigma}=\frac{1}{2}(\gamma_v+\gamma_{k_0})=0.299/N$ and $T_0=v k_0$.
\subsubsection{Heat capacity and Entropy}
The scaling of the free energy density is:
\begin{equation}
	f(T,\mu)= b^{-1} Z_x^{-1} Z_{\omega}^{-1} ~f(Z_{\omega} T, Z_{\omega} \mu)
	\:,
\end{equation}
which reproduces the above scaling dimension for the particle density via $n=\frac{\partial f}{\partial \mu}$.
For the heat capacity follows
\begin{eqnarray}
	C(T) &=& b^{-1} Z_x^{-1} ~C(Z_{\omega} T) \nonumber \\
	&=& b^{-(1+\phi)} C(b^{z} T) \nonumber \\
	&\propto& T^{\frac{1+\phi}{z}} \propto T^{\phi_C}
\:,
\end{eqnarray}
with
\begin{equation}
	\phi_C
	=
	\frac{3}{2} -\frac{1}{2} \left( \gamma_{k_0}-3 \gamma_v \right) = \frac{3}{2} +\frac{0.4255}{N}
	\:.
\end{equation}
The entropy is defined via the capacity by 
\begin{equation}
	C= T \frac{\partial S}{\partial T}
	\:.
\end{equation}
We thus find for the entropy the same scaling dimension as for the heat capacity
\begin{equation}
    S \propto T^{\frac{1+\phi}{z}} \propto T^{\phi_C} \propto T^{\frac{3}{2} +\frac{0.4255}{N}}
    \:.
\end{equation}

\subsubsection{Shear viscosity}
Here, the scaling behavior of the viscosity is studied. The shear viscosity $\eta_{\alpha \beta \gamma \delta}$ is also defined by the Kubo-formula~\cite{PhysRevB.86.245309}:
\begin{equation}
		\eta_{\alpha \beta \gamma \delta} \propto \frac{1}{\omega} \im \av{[\tau_{\alpha \beta}, \tau_{\gamma \delta}]}
		\:,
\end{equation}
where $\im \av{[\tau_{\alpha \beta} ,\tau_{\gamma \delta}]}$ is the correlation function of the energy-stress tensor. The Kubo-formula defining the shear viscosity can also be expressed by the strain generator
$\mathcal{J}_{\alpha \beta}= \mathcal{L}_{\alpha \beta}+\mathcal{S}_{\alpha \beta}=x_\alpha p_\beta + \frac{i}{2} \delta_{\alpha \beta}-\epsilon_{\alpha \beta \gamma }\frac{\sigma_{\gamma}}{4}$ and has the form~\cite{PhysRevB.86.245309}:
\begin{equation}
 \eta_{\alpha \beta \gamma \delta} \propto ~\omega~ \im \av{[\mathcal{J}_{\alpha \beta}, \mathcal{J}_{\gamma \delta}]}
 \:.
\end{equation}
%
Upon assuming for the operator $\mathcal{L}_{\alpha \beta}=x_\alpha p_\beta + \frac{i}{2} \delta_{\alpha \beta}$ the same dimensionality as the particle density times the momentum and the corresponding
spatial coordinate, which has the dimensionality of the inverse momentum, we find
\begin{eqnarray}
 Z_{\mathcal{L}_{\alpha \alpha}} &=& Z_n\\
 Z_{\mathcal{L}_{xy}} &=& \frac{Z_x}{b} Z_n= b^{\phi-1} Z_n\\
 Z_{\mathcal{L}_{yx}} &=& \frac{b}{Z_x} Z_n= b^{-(\phi-1)} Z_n
 \:.
\end{eqnarray}
%
The viscosity coefficients have thus the scaling dimension
 \begin{eqnarray}
	Z_{\eta_{ijkl}} &=&  Z_x^{-1} b^{-1}= b^{-(\phi+1)}\\
	Z_{\eta_{xyxy}} &=&  Z_x b^{-3}=b^{\phi-3}\\
	Z_{\eta_{yxyx}} &=& Z_x^{-3} b = b^{-3 \phi+1}
	\:,
\end{eqnarray}
where the Roman indices denote all other viscosity coefficients but $\eta_{xyxy}$ and $\eta_{yxyx}$.
The scaling implies
\begin{eqnarray}
	\eta_{ijkl}(T) &=& b^{-(\phi+1)}~	\eta_{s}(b^{z} T)\\
	\eta_{xyxy}(T) &=& b^{\phi-3} ~\eta_{xyxyx}(b^{z} T)\\
	\eta_{yxyx}(T) &=& b^{-3 \phi+1} ~	\eta_{yxyx}(b^{z} T)
	\:.
\end{eqnarray}
It follows
\begin{eqnarray}
	\eta_{ijkl} &\propto & \left( \frac{T}{T_0} \right)^{\frac{\phi+1}{z}} \frac{k^2_0}{\hbar}~\mathcal{C}_s \nonumber \\
	& \propto & \left( \frac{T}{T_0} \right)^{3/2+\frac{0.4255}{N}} \frac{k^2_0}{\hbar}~\mathcal{C}_s\\
 	\eta_{xyxy} & \propto & \left( \frac{T}{T_0} \right)^{-\frac{\phi-3}{z}} \frac{k^2_0}{\hbar}~\mathcal{C}_{xyxy} \nonumber \\
	       & \propto & \left( \frac{T}{T_0} \right)^{5/2+\frac{1.02445}{N}} \frac{k^2_0}{\hbar}~\mathcal{C}_{xyxy} \:,  \\
	\eta_{yxyx} & \propto & \left( \frac{T}{T_0} \right)^{-\frac{(-3 \phi+1)}{z}} \frac{k^2_0}{\hbar} ~\mathcal{C}_{yxyx} \nonumber \\ 
	       & \propto &\left( \frac{T}{T_0} \right)^{1/2-\frac{0.17335}{N}} \frac{k^2_0}{\hbar} ~\mathcal{C}_{yxyx} 
	\:.
\end{eqnarray}
%
%


\begin{thebibliography}{46}%
\makeatletter
\providecommand \@ifxundefined [1]{%
 \@ifx{#1\undefined}
}%
\providecommand \@ifnum [1]{%
 \ifnum #1\expandafter \@firstoftwo
 \else \expandafter \@secondoftwo
 \fi
}%
\providecommand \@ifx [1]{%
 \ifx #1\expandafter \@firstoftwo
 \else \expandafter \@secondoftwo
 \fi
}%
\providecommand \natexlab [1]{#1}%
\providecommand \enquote  [1]{``#1''}%
\providecommand \bibnamefont  [1]{#1}%
\providecommand \bibfnamefont [1]{#1}%
\providecommand \citenamefont [1]{#1}%
\providecommand \href@noop [0]{\@secondoftwo}%
\providecommand \href [0]{\begingroup \@sanitize@url \@href}%
\providecommand \@href[1]{\@@startlink{#1}\@@href}%
\providecommand \@@href[1]{\endgroup#1\@@endlink}%
\providecommand \@sanitize@url [0]{\catcode `\\12\catcode `\$12\catcode
  `\&12\catcode `\#12\catcode `\^12\catcode `\_12\catcode `\%12\relax}%
\providecommand \@@startlink[1]{}%
\providecommand \@@endlink[0]{}%
\providecommand \url  [0]{\begingroup\@sanitize@url \@url }%
\providecommand \@url [1]{\endgroup\@href {#1}{\urlprefix }}%
\providecommand \urlprefix  [0]{URL }%
\providecommand \Eprint [0]{\href }%
\providecommand \doibase [0]{http://dx.doi.org/}%
\providecommand \selectlanguage [0]{\@gobble}%
\providecommand \bibinfo  [0]{\@secondoftwo}%
\providecommand \bibfield  [0]{\@secondoftwo}%
\providecommand \translation [1]{[#1]}%
\providecommand \BibitemOpen [0]{}%
\providecommand \bibitemStop [0]{}%
\providecommand \bibitemNoStop [0]{.\EOS\space}%
\providecommand \EOS [0]{\spacefactor3000\relax}%
\providecommand \BibitemShut  [1]{\csname bibitem#1\endcsname}%
\let\auto@bib@innerbib\@empty
\bibitem [{\citenamefont {Vollhardt}\ and\ \citenamefont
  {W\"olfle}(1990)}]{The-Superfluid-Phases-of-Helium3}%
  \BibitemOpen
  \bibfield  {author} {\bibinfo {author} {\bibfnamefont {D.}~\bibnamefont
  {Vollhardt}}\ and\ \bibinfo {author} {\bibfnamefont {P.}~\bibnamefont
  {W\"olfle}},\ }\href@noop {} {\emph {\bibinfo {title} {The Superfluid Phases
  of Helium 3}}}\ (\bibinfo  {publisher} {Taylor \& Francis Ltd},\ \bibinfo
  {year} {1990})\BibitemShut {NoStop}%
\bibitem [{\citenamefont {de~Jong}\ and\ \citenamefont
  {Molenkamp}(1995)}]{PhysRevB.51.13389}%
  \BibitemOpen
  \bibfield  {author} {\bibinfo {author} {\bibfnamefont {M.~J.~M.}\
  \bibnamefont {de~Jong}}\ and\ \bibinfo {author} {\bibfnamefont {L.~W.}\
  \bibnamefont {Molenkamp}},\ }\href {\doibase 10.1103/PhysRevB.51.13389}
  {\bibfield  {journal} {\bibinfo  {journal} {Phys. Rev. B}\ }\textbf {\bibinfo
  {volume} {51}},\ \bibinfo {pages} {13389} (\bibinfo {year}
  {1995})}\BibitemShut {NoStop}%
\bibitem [{\citenamefont {Chafin}\ and\ \citenamefont
  {Sch\"afer}(2013)}]{PhysRevA.87.023629}%
  \BibitemOpen
  \bibfield  {author} {\bibinfo {author} {\bibfnamefont {C.}~\bibnamefont
  {Chafin}}\ and\ \bibinfo {author} {\bibfnamefont {T.}~\bibnamefont
  {Sch\"afer}},\ }\href {\doibase 10.1103/PhysRevA.87.023629} {\bibfield
  {journal} {\bibinfo  {journal} {Phys. Rev. A}\ }\textbf {\bibinfo {volume}
  {87}},\ \bibinfo {pages} {023629} (\bibinfo {year} {2013})}\BibitemShut
  {NoStop}%
\bibitem [{\citenamefont {Enss}\ \emph {et~al.}(2011)\citenamefont {Enss},
  \citenamefont {Haussmann},\ and\ \citenamefont {Zwerger}}]{ENSS2011770}%
  \BibitemOpen
  \bibfield  {author} {\bibinfo {author} {\bibfnamefont {T.}~\bibnamefont
  {Enss}}, \bibinfo {author} {\bibfnamefont {R.}~\bibnamefont {Haussmann}}, \
  and\ \bibinfo {author} {\bibfnamefont {W.}~\bibnamefont {Zwerger}},\ }\href
  {\doibase 10.1016/j.aop.2010.10.002} {\bibfield  {journal} {\bibinfo
  {journal} {Ann. Phys. (Amsterdam)}\ }\textbf {\bibinfo {volume} {326}},\
  \bibinfo {pages} {770 } (\bibinfo {year} {2011})}\BibitemShut {NoStop}%
\bibitem [{\citenamefont
  {Sch\"afer}(2014)}]{Fluid-Dynamics-and-Viscosity-in-Strongly-Correlated-Fluids}%
  \BibitemOpen
  \bibfield  {author} {\bibinfo {author} {\bibfnamefont {T.}~\bibnamefont
  {Sch\"afer}},\ }\href {\doibase 10.1146/annurev-nucl-102313-025439}
  {\bibfield  {journal} {\bibinfo  {journal} {Annu. Rev. Nucl. Part. Sci.}\
  }\textbf {\bibinfo {volume} {64}},\ \bibinfo {pages} {125} (\bibinfo {year}
  {2014})}\BibitemShut {NoStop}%
\bibitem [{\citenamefont {Shuryak}(2004)}]{SHURYAK2004273}%
  \BibitemOpen
  \bibfield  {author} {\bibinfo {author} {\bibfnamefont {E.}~\bibnamefont
  {Shuryak}},\ }\href {\doibase 10.1016/j.ppnp.2004.02.025} {\bibfield
  {journal} {\bibinfo  {journal} {Prog. Part. Nucl. Phys.}\ }\textbf {\bibinfo
  {volume} {53}},\ \bibinfo {pages} {273 } (\bibinfo {year}
  {2004})}\BibitemShut {NoStop}%
\bibitem [{\citenamefont {Gurzhi}(1963)}]{Gurzhi}%
  \BibitemOpen
  \bibfield  {author} {\bibinfo {author} {\bibfnamefont {R.~N.}\ \bibnamefont
  {Gurzhi}},\ }\href@noop {} {\bibfield  {journal} {\bibinfo  {journal} {Zh.
  Eksp. Teor. Fiz.}\ }\textbf {\bibinfo {volume} {44}},\ \bibinfo {pages} {771}
  (\bibinfo {year} {1963})},\ \bibinfo {note} {[Sov. Phys. JETP 17, 521
  (1963)]}\BibitemShut {NoStop}%
\bibitem [{\citenamefont {Gurzhi}(1968)}]{GurzhiUFN}%
  \BibitemOpen
  \bibfield  {author} {\bibinfo {author} {\bibfnamefont {R.~N.}\ \bibnamefont
  {Gurzhi}},\ }\href@noop {} {\bibfield  {journal} {\bibinfo  {journal} {Usp.
  Fiz. Nauk}\ }\textbf {\bibinfo {volume} {94}},\ \bibinfo {pages} {689}
  (\bibinfo {year} {1968})},\ \bibinfo {note} {[Sov. Phys. Usp. 11, 255
  (1968)]}\BibitemShut {NoStop}%
\bibitem [{\citenamefont {Crossno}\ \emph {et~al.}(2016)\citenamefont
  {Crossno}, \citenamefont {Shi}, \citenamefont {Wang}, \citenamefont {Liu},
  \citenamefont {Harzheim}, \citenamefont {Lucas}, \citenamefont {Sachdev},
  \citenamefont {Kim}, \citenamefont {Taniguchi}, \citenamefont {Watanabe},
  \citenamefont {Ohki},\ and\ \citenamefont {Fong}}]{Science351.1058}%
  \BibitemOpen
  \bibfield  {author} {\bibinfo {author} {\bibfnamefont {J.}~\bibnamefont
  {Crossno}}, \bibinfo {author} {\bibfnamefont {J.~K.}\ \bibnamefont {Shi}},
  \bibinfo {author} {\bibfnamefont {K.}~\bibnamefont {Wang}}, \bibinfo {author}
  {\bibfnamefont {X.}~\bibnamefont {Liu}}, \bibinfo {author} {\bibfnamefont
  {A.}~\bibnamefont {Harzheim}}, \bibinfo {author} {\bibfnamefont
  {A.}~\bibnamefont {Lucas}}, \bibinfo {author} {\bibfnamefont
  {S.}~\bibnamefont {Sachdev}}, \bibinfo {author} {\bibfnamefont
  {P.}~\bibnamefont {Kim}}, \bibinfo {author} {\bibfnamefont {T.}~\bibnamefont
  {Taniguchi}}, \bibinfo {author} {\bibfnamefont {K.}~\bibnamefont {Watanabe}},
  \bibinfo {author} {\bibfnamefont {T.~A.}\ \bibnamefont {Ohki}}, \ and\
  \bibinfo {author} {\bibfnamefont {K.~C.}\ \bibnamefont {Fong}},\ }\href
  {\doibase 10.1126/science.aad0343} {\bibfield  {journal} {\bibinfo  {journal}
  {Science}\ }\textbf {\bibinfo {volume} {351}},\ \bibinfo {pages} {1058}
  (\bibinfo {year} {2016})}\BibitemShut {NoStop}%
\bibitem [{\citenamefont {Guo}\ \emph {et~al.}(2017)\citenamefont {Guo},
  \citenamefont {Ilseven}, \citenamefont {Falkovich},\ and\ \citenamefont
  {Levitov}}]{Falkovich}%
  \BibitemOpen
  \bibfield  {author} {\bibinfo {author} {\bibfnamefont {H.}~\bibnamefont
  {Guo}}, \bibinfo {author} {\bibfnamefont {E.}~\bibnamefont {Ilseven}},
  \bibinfo {author} {\bibfnamefont {G.}~\bibnamefont {Falkovich}}, \ and\
  \bibinfo {author} {\bibfnamefont {L.~S.}\ \bibnamefont {Levitov}},\ }\href
  {\doibase 10.1073/pnas.1612181114} {\bibfield  {journal} {\bibinfo  {journal}
  {Proc. Natl. Acad. Sci. U.S.A.}\ }\textbf {\bibinfo {volume} {114}},\ \bibinfo {pages} {3068}
  (\bibinfo {year} {2017})}\BibitemShut {NoStop}%
\bibitem [{\citenamefont {{Krishna Kumar}}\ \emph {et~al.}(2017)\citenamefont
  {{Krishna Kumar}}, \citenamefont {Bandurin}, \citenamefont {Pellegrino},
  \citenamefont {Cao}, \citenamefont {Principi}, \citenamefont {Guo},
  \citenamefont {Auton}, \citenamefont {Shalom}, \citenamefont {Ponomarenko},
  \citenamefont {Falkovich}, \citenamefont {Grigorieva}, \citenamefont
  {Levitov}, \citenamefont {Polini},\ and\ \citenamefont {Geim}}]{Geim17}%
  \BibitemOpen
  \bibfield  {author} {\bibinfo {author} {\bibfnamefont {R.}~\bibnamefont
  {{Krishna Kumar}}}, \bibinfo {author} {\bibfnamefont {D.~A.}\ \bibnamefont
  {Bandurin}}, \bibinfo {author} {\bibfnamefont {F.~M.~D.}\ \bibnamefont
  {Pellegrino}}, \bibinfo {author} {\bibfnamefont {Y.}~\bibnamefont {Cao}},
  \bibinfo {author} {\bibfnamefont {A.}~\bibnamefont {Principi}}, \bibinfo
  {author} {\bibfnamefont {H.}~\bibnamefont {Guo}}, \bibinfo {author}
  {\bibfnamefont {G.~H.}\ \bibnamefont {Auton}}, \bibinfo {author}
  {\bibfnamefont {M.~B.}\ \bibnamefont {Shalom}}, \bibinfo {author}
  {\bibfnamefont {L.~A.}\ \bibnamefont {Ponomarenko}}, \bibinfo {author}
  {\bibfnamefont {G.}~\bibnamefont {Falkovich}}, \bibinfo {author}
  {\bibfnamefont {I.~V.}\ \bibnamefont {Grigorieva}}, \bibinfo {author}
  {\bibfnamefont {L.~S.}\ \bibnamefont {Levitov}}, \bibinfo {author}
  {\bibfnamefont {M.}~\bibnamefont {Polini}}, \ and\ \bibinfo {author}
  {\bibfnamefont {A.~K.}\ \bibnamefont {Geim}},\ }\href {\doibase
  10.1038/nphys4240} {\bibfield  {journal} {\bibinfo  {journal} {Nat.
  Phys.}\ }\textbf {\bibinfo {volume} {13}},\ \bibinfo {pages} {1182}
  (\bibinfo {year} {2017})}\BibitemShut {NoStop}%
\bibitem [{\citenamefont {Bandurin}\ \emph {et~al.}(2016)\citenamefont
  {Bandurin}, \citenamefont {Torre}, \citenamefont {Krishna~Kumar},
  \citenamefont {Ben~Shalom}, \citenamefont {Tomadin}, \citenamefont
  {Principi}, \citenamefont {Auton}, \citenamefont {Khestanova1}, \citenamefont
  {Novoselov}, \citenamefont {Grigorieva1}, \citenamefont {Ponomarenko},
  \citenamefont {Geim},\ and\ \citenamefont {Polini}}]{Science351.1055}%
  \BibitemOpen
  \bibfield  {author} {\bibinfo {author} {\bibfnamefont {D.~A.}\ \bibnamefont
  {Bandurin}}, \bibinfo {author} {\bibfnamefont {I.}~\bibnamefont {Torre}},
  \bibinfo {author} {\bibfnamefont {R.}~\bibnamefont {Krishna~Kumar}}, \bibinfo
  {author} {\bibfnamefont {M.}~\bibnamefont {Ben~Shalom}}, \bibinfo {author}
  {\bibfnamefont {A.}~\bibnamefont {Tomadin}}, \bibinfo {author} {\bibfnamefont
  {A.}~\bibnamefont {Principi}}, \bibinfo {author} {\bibfnamefont {G.~H.}\
  \bibnamefont {Auton}}, \bibinfo {author} {\bibfnamefont {E.}~\bibnamefont
  {Khestanova1}}, \bibinfo {author} {\bibfnamefont {K.~S.}\ \bibnamefont
  {Novoselov}}, \bibinfo {author} {\bibfnamefont {I.~V.}\ \bibnamefont
  {Grigorieva1}}, \bibinfo {author} {\bibfnamefont {L.~A.}\ \bibnamefont
  {Ponomarenko}}, \bibinfo {author} {\bibfnamefont {A.~K.}\ \bibnamefont
  {Geim}}, \ and\ \bibinfo {author} {\bibfnamefont {M.}~\bibnamefont
  {Polini}},\ }\href {\doibase 10.1126/science.aad0201} {\bibfield  {journal}
  {\bibinfo  {journal} {Science}\ }\textbf {\bibinfo {volume} {351}},\ \bibinfo
  {pages} {1055} (\bibinfo {year} {2016})}\BibitemShut {NoStop}%
\bibitem [{\citenamefont {Levitov}\ and\ \citenamefont
  {Falkovich}(2016)}]{Levitov2016}%
  \BibitemOpen
  \bibfield  {author} {\bibinfo {author} {\bibfnamefont {L.~S.}\ \bibnamefont
  {Levitov}}\ and\ \bibinfo {author} {\bibfnamefont {G.}~\bibnamefont
  {Falkovich}},\ }\href {\doibase 10.1038/nphys3667} {\bibfield  {journal}
  {\bibinfo  {journal} {Nat. Phys.}\ }\textbf {\bibinfo {volume} {12}},\
  \bibinfo {pages} {672} (\bibinfo {year} {2016})}\BibitemShut {NoStop}%
\bibitem [{\citenamefont {Titov}\ \emph {et~al.}(2013)\citenamefont {Titov},
  \citenamefont {Gorbachev}, \citenamefont {Narozhny}, \citenamefont
  {Tudorovskiy}, \citenamefont {Sch\"utt}, \citenamefont {Ostrovsky},
  \citenamefont {Gornyi}, \citenamefont {Mirlin}, \citenamefont {Katsnelson},
  \citenamefont {Novoselov}, \citenamefont {Geim},\ and\ \citenamefont
  {Ponomarenko}}]{PhysRevLett.111.166601}%
  \BibitemOpen
  \bibfield  {author} {\bibinfo {author} {\bibfnamefont {M.}~\bibnamefont
  {Titov}}, \bibinfo {author} {\bibfnamefont {R.~V.}\ \bibnamefont
  {Gorbachev}}, \bibinfo {author} {\bibfnamefont {B.~N.}\ \bibnamefont
  {Narozhny}}, \bibinfo {author} {\bibfnamefont {T.}~\bibnamefont
  {Tudorovskiy}}, \bibinfo {author} {\bibfnamefont {M.}~\bibnamefont
  {Sch\"utt}}, \bibinfo {author} {\bibfnamefont {P.~M.}\ \bibnamefont
  {Ostrovsky}}, \bibinfo {author} {\bibfnamefont {I.~V.}\ \bibnamefont
  {Gornyi}}, \bibinfo {author} {\bibfnamefont {A.~D.}\ \bibnamefont {Mirlin}},
  \bibinfo {author} {\bibfnamefont {M.~I.}\ \bibnamefont {Katsnelson}},
  \bibinfo {author} {\bibfnamefont {K.~S.}\ \bibnamefont {Novoselov}}, \bibinfo
  {author} {\bibfnamefont {A.~K.}\ \bibnamefont {Geim}}, \ and\ \bibinfo
  {author} {\bibfnamefont {L.~A.}\ \bibnamefont {Ponomarenko}},\ }\href
  {\doibase 10.1103/PhysRevLett.111.166601} {\bibfield  {journal} {\bibinfo
  {journal} {Phys. Rev. Lett.}\ }\textbf {\bibinfo {volume} {111}},\ \bibinfo
  {pages} {166601} (\bibinfo {year} {2013})}\BibitemShut {NoStop}%
\bibitem [{\citenamefont {Narozhny}\ \emph {et~al.}(2017)\citenamefont
  {Narozhny}, \citenamefont {Gornyi}, \citenamefont {Mirlin},\ and\
  \citenamefont {Schmalian}}]{ANDP:ANDP201700043}%
  \BibitemOpen
  \bibfield  {author} {\bibinfo {author} {\bibfnamefont {B.~N.}\ \bibnamefont
  {Narozhny}}, \bibinfo {author} {\bibfnamefont {I.~V.}\ \bibnamefont
  {Gornyi}}, \bibinfo {author} {\bibfnamefont {A.~D.}\ \bibnamefont {Mirlin}},
  \ and\ \bibinfo {author} {\bibfnamefont {J.}~\bibnamefont {Schmalian}},\
  }\href {\doibase 10.1002/andp.201700043} {\bibfield  {journal} {\bibinfo
  {journal} {Annalen der Physik}\ }\textbf {\bibinfo {volume} {529}},\ \bibinfo
  {pages} {1700043} (\bibinfo {year} {2017})}\BibitemShut {NoStop}%
\bibitem [{\citenamefont {Lucas}\ and\ \citenamefont
  {Fong}(2018)}]{LucasFong2018}%
  \BibitemOpen
  \bibfield  {author} {\bibinfo {author} {\bibfnamefont {A.}~\bibnamefont
  {Lucas}}\ and\ \bibinfo {author} {\bibfnamefont {K.~C.}\ \bibnamefont
  {Fong}},\ }\href{http://stacks.iop.org/0953-8984/30/i=5/a=053001} {\bibfield  {journal} {\bibinfo  {journal} {J.
  Phys. Cond. Matter}\ }\textbf {\bibinfo {volume} {30}},\ \bibinfo
  {pages} {053001} (\bibinfo {year} {2018})}\BibitemShut {NoStop}%
\bibitem [{\citenamefont {Moll}\ \emph {et~al.}(2016)\citenamefont {Moll},
  \citenamefont {Kushwaha}, \citenamefont {Nandi}, \citenamefont {Schmidt},\
  and\ \citenamefont {Mackenzie}}]{Science351.1061}%
  \BibitemOpen
  \bibfield  {author} {\bibinfo {author} {\bibfnamefont {P.~J.~W.}\
  \bibnamefont {Moll}}, \bibinfo {author} {\bibfnamefont {P.}~\bibnamefont
  {Kushwaha}}, \bibinfo {author} {\bibfnamefont {N.}~\bibnamefont {Nandi}},
  \bibinfo {author} {\bibfnamefont {B.}~\bibnamefont {Schmidt}}, \ and\
  \bibinfo {author} {\bibfnamefont {A.~P.}\ \bibnamefont {Mackenzie}},\ }\href
  {\doibase 10.1126/science.aac8385} {\bibfield  {journal} {\bibinfo  {journal}
  {Science}\ }\textbf {\bibinfo {volume} {351}},\ \bibinfo {pages} {1061}
  (\bibinfo {year} {2016})}\BibitemShut {NoStop}%
\bibitem [{\citenamefont {{Gooth}}\ \emph {et~al.}(2017)\citenamefont
  {{Gooth}}, \citenamefont {{Menges}}, \citenamefont {{Shekhar}}, \citenamefont
  {{S{\"u}{\ss}}}, \citenamefont {{Kumar}}, \citenamefont {{Sun}},
  \citenamefont {{Drechsler}}, \citenamefont {{Zierold}}, \citenamefont
  {{Felser}},\ and\ \citenamefont {{Gotsmann}}}]{2017arXiv170605925G}%
  \BibitemOpen
  \bibfield  {author} {\bibinfo {author} {\bibfnamefont {J.}~\bibnamefont
  {{Gooth}}}, \bibinfo {author} {\bibfnamefont {F.}~\bibnamefont {{Menges}}},
  \bibinfo {author} {\bibfnamefont {C.}~\bibnamefont {{Shekhar}}}, \bibinfo
  {author} {\bibfnamefont {V.}~\bibnamefont {{S{\"u}{\ss}}}}, \bibinfo {author}
  {\bibfnamefont {N.}~\bibnamefont {{Kumar}}}, \bibinfo {author} {\bibfnamefont
  {Y.}~\bibnamefont {{Sun}}}, \bibinfo {author} {\bibfnamefont
  {U.}~\bibnamefont {{Drechsler}}}, \bibinfo {author} {\bibfnamefont
  {R.}~\bibnamefont {{Zierold}}}, \bibinfo {author} {\bibfnamefont
  {C.}~\bibnamefont {{Felser}}}, \ and\ \bibinfo {author} {\bibfnamefont
  {B.}~\bibnamefont {{Gotsmann}}},\ }\href@noop {} {\  (\bibinfo {year}
  {2017})},\ \bibinfo {note} {{arXiv:1706.05925}}\BibitemShut {NoStop}%
\bibitem [{\citenamefont {{Maldacena}}(1998)}]{1998AdTMP...2..231M}%
  \BibitemOpen
  \bibfield  {author} {\bibinfo {author} {\bibfnamefont {J.~M.}\ \bibnamefont
  {{Maldacena}}},\ }\href {\doibase 10.4310/ATMP.1998.v2.n2.a1} {\bibfield
  {journal} {\bibinfo  {journal} {Adv. Theor. Math. Phys.}\ }\textbf {\bibinfo
  {volume} {2}},\ \bibinfo {pages} {231} (\bibinfo {year} {1998})}\BibitemShut
  {NoStop}%
\bibitem [{\citenamefont {Kovtun}\ \emph {et~al.}(2005)\citenamefont {Kovtun},
  \citenamefont {Son},\ and\ \citenamefont
  {Starinets}}]{PhysRevLett.94.111601}%
  \BibitemOpen
  \bibfield  {author} {\bibinfo {author} {\bibfnamefont {P.~K.}\ \bibnamefont
  {Kovtun}}, \bibinfo {author} {\bibfnamefont {D.~T.}\ \bibnamefont {Son}}, \
  and\ \bibinfo {author} {\bibfnamefont {A.~O.}\ \bibnamefont {Starinets}},\
  }\href {\doibase 10.1103/PhysRevLett.94.111601} {\bibfield  {journal}
  {\bibinfo  {journal} {Phys. Rev. Lett.}\ }\textbf {\bibinfo {volume} {94}},\
  \bibinfo {pages} {111601} (\bibinfo {year} {2005})}\BibitemShut {NoStop}%
\bibitem [{\citenamefont {Lifshitz}\ and\ \citenamefont
  {Pitaevskii}(1981)}]{dau10}%
  \BibitemOpen
  \bibfield  {author} {\bibinfo {author} {\bibfnamefont {E.~M.}\ \bibnamefont
  {Lifshitz}}\ and\ \bibinfo {author} {\bibfnamefont {L.~P.}\ \bibnamefont
  {Pitaevskii}},\ }\href@noop {} {\emph {\bibinfo {title} {Physical
  Kinetics}}}\ (\bibinfo  {publisher} {Pergamon Press (Oxford)},\ \bibinfo
  {year} {1981})\BibitemShut {NoStop}%
\bibitem [{\citenamefont {Rebhan}\ and\ \citenamefont
  {Steineder}(2012)}]{PhysRevLett.108.021601}%
  \BibitemOpen
  \bibfield  {author} {\bibinfo {author} {\bibfnamefont {A.}~\bibnamefont
  {Rebhan}}\ and\ \bibinfo {author} {\bibfnamefont {D.}~\bibnamefont
  {Steineder}},\ }\href {\doibase 10.1103/PhysRevLett.108.021601} {\bibfield
  {journal} {\bibinfo  {journal} {Phys. Rev. Lett.}\ }\textbf {\bibinfo
  {volume} {108}},\ \bibinfo {pages} {021601} (\bibinfo {year}
  {2012})}\BibitemShut {NoStop}%
\bibitem [{\citenamefont {Jain}\ \emph {et~al.}(2015)\citenamefont {Jain},
  \citenamefont {Samanta},\ and\ \citenamefont {Trivedi}}]{Jain2015}%
  \BibitemOpen
  \bibfield  {author} {\bibinfo {author} {\bibfnamefont {S.}~\bibnamefont
  {Jain}}, \bibinfo {author} {\bibfnamefont {R.}~\bibnamefont {Samanta}}, \
  and\ \bibinfo {author} {\bibfnamefont {S.~P.}\ \bibnamefont {Trivedi}},\
  }\href {\doibase 10.1007/JHEP10(2015)028} {\bibfield  {journal} {\bibinfo
  {journal} {J. High Energy Phys.}\ }\textbf {\bibinfo {volume}
  {28}},\ \bibinfo {pages} {} (\bibinfo {year} {2015})}\BibitemShut
  {NoStop}%
\bibitem [{\citenamefont {Cremonini}(2011)}]{doi:10.1142/S0217984911027315}%
  \BibitemOpen
  \bibfield  {author} {\bibinfo {author} {\bibfnamefont {S.}~\bibnamefont
  {Cremonini}},\ }\href {\doibase 10.1142/S0217984911027315} {\bibfield
  {journal} {\bibinfo  {journal} {Mod. Phys. Lett. B}\ }\textbf {\bibinfo
  {volume} {25}},\ \bibinfo {pages} {1867} (\bibinfo {year}
  {2011})}\BibitemShut {NoStop}%
\bibitem [{\citenamefont {{Samanta}}\ \emph {et~al.}(2016)\citenamefont 
  {{Samanta}}, \citenamefont {{Sharma}},\ and\ \citenamefont
  {{Trivedi}}}]{2016arXiv160704799S}%
  \BibitemOpen
  \bibfield  {author} {\bibinfo {author} {\bibfnamefont {R.}~\bibnamefont
  {{Samanta}}}, \bibinfo {author} {\bibfnamefont {R.}~\bibnamefont {{Sharma}}},
  \ and\ \bibinfo {author} {\bibfnamefont {S.~P.}\ \bibnamefont {{Trivedi}}},\
  }
  \href {\doibase
  10.1103/PhysRevA.96.053601} {\bibfield  {journal} {\bibinfo  {journal}
  {Phys. Rev. A}\ }\textbf {\bibinfo {volume} {96}},\ \bibinfo {pages}
  {053601} (\bibinfo {year} {2017})}\BibitemShut {NoStop}
\bibitem [{\citenamefont {Isobe}\ \emph {et~al.}(2016)\citenamefont {Isobe},
  \citenamefont {Yang}, \citenamefont {Chubukov}, \citenamefont {Schmalian},\
  and\ \citenamefont {Nagaosa}}]{PhysRevLett.116.076803}%
  \BibitemOpen
  \bibfield  {author} {\bibinfo {author} {\bibfnamefont {H.}~\bibnamefont
  {Isobe}}, \bibinfo {author} {\bibfnamefont {B.-J.}\ \bibnamefont {Yang}},
  \bibinfo {author} {\bibfnamefont {A.}~\bibnamefont {Chubukov}}, \bibinfo
  {author} {\bibfnamefont {J.}~\bibnamefont {Schmalian}}, \ and\ \bibinfo
  {author} {\bibfnamefont {N.}~\bibnamefont {Nagaosa}},\ }\href {\doibase
  10.1103/PhysRevLett.116.076803} {\bibfield  {journal} {\bibinfo  {journal}
  {Phys. Rev. Lett.}\ }\textbf {\bibinfo {volume} {116}},\ \bibinfo {pages}
  {076803} (\bibinfo {year} {2016})}\BibitemShut {NoStop}%
\bibitem [{\citenamefont {Kobayashi}\ \emph {et~al.}(2011)\citenamefont
  {Kobayashi}, \citenamefont {Suzumura}, \citenamefont {Pi\'echon},\ and\
  \citenamefont {Montambaux}}]{PhysRevB.84.075450}%
  \BibitemOpen
  \bibfield  {author} {\bibinfo {author} {\bibfnamefont {A.}~\bibnamefont
  {Kobayashi}}, \bibinfo {author} {\bibfnamefont {Y.}~\bibnamefont {Suzumura}},
  \bibinfo {author} {\bibfnamefont {F.}~\bibnamefont {Pi\'echon}}, \ and\
  \bibinfo {author} {\bibfnamefont {G.}~\bibnamefont {Montambaux}},\ }\href
  {\doibase 10.1103/PhysRevB.84.075450} {\bibfield  {journal} {\bibinfo
  {journal} {Phys. Rev. B}\ }\textbf {\bibinfo {volume} {84}},\ \bibinfo
  {pages} {075450} (\bibinfo {year} {2011})}\BibitemShut {NoStop}%
\bibitem [{\citenamefont {Pardo}\ and\ \citenamefont
  {Pickett}(2009)}]{PhysRevLett.102.166803}%
  \BibitemOpen
  \bibfield  {author} {\bibinfo {author} {\bibfnamefont {V.}~\bibnamefont
  {Pardo}}\ and\ \bibinfo {author} {\bibfnamefont {W.~E.}\ \bibnamefont
  {Pickett}},\ }\href {\doibase 10.1103/PhysRevLett.102.166803} {\bibfield
  {journal} {\bibinfo  {journal} {Phys. Rev. Lett.}\ }\textbf {\bibinfo
  {volume} {102}},\ \bibinfo {pages} {166803} (\bibinfo {year}
  {2009})}\BibitemShut {NoStop}%
\bibitem [{\citenamefont {Banerjee}\ \emph {et~al.}(2009)\citenamefont
  {Banerjee}, \citenamefont {Singh}, \citenamefont {Pardo},\ and\ \citenamefont
  {Pickett}}]{PhysRevLett.103.016402}%
  \BibitemOpen
  \bibfield  {author} {\bibinfo {author} {\bibfnamefont {S.}~\bibnamefont
  {Banerjee}}, \bibinfo {author} {\bibfnamefont {R.~R.~P.}\ \bibnamefont
  {Singh}}, \bibinfo {author} {\bibfnamefont {V.}~\bibnamefont {Pardo}}, \ and\
  \bibinfo {author} {\bibfnamefont {W.~E.}\ \bibnamefont {Pickett}},\ }\href
  {\doibase 10.1103/PhysRevLett.103.016402} {\bibfield  {journal} {\bibinfo
  {journal} {Phys. Rev. Lett.}\ }\textbf {\bibinfo {volume} {103}},\ \bibinfo
  {pages} {016402} (\bibinfo {year} {2009})}\BibitemShut {NoStop}%
\bibitem [{\citenamefont {Fang}\ and\ \citenamefont {Fu}(2015)}]{Fang2015}%
  \BibitemOpen
  \bibfield  {author} {\bibinfo {author} {\bibfnamefont {C.}~\bibnamefont
  {Fang}}\ and\ \bibinfo {author} {\bibfnamefont {L.}~\bibnamefont {Fu}},\
  }\href {\doibase 10.1103/PhysRevB.91.161105} {\bibfield  {journal} {\bibinfo
  {journal} {Phys. Rev. B}\ }\textbf {\bibinfo {volume} {91}},\ \bibinfo
  {pages} {161105} (\bibinfo {year} {2015})}\BibitemShut {NoStop}%
\bibitem [{\citenamefont {Huang}\ \emph {et~al.}(2016)\citenamefont {Huang},
  \citenamefont {Xu}, \citenamefont {Belopolski}, \citenamefont {Lee},
  \citenamefont {Chang}, \citenamefont {Chang}, \citenamefont {Wang},
  \citenamefont {Alidoust}, \citenamefont {Bian}, \citenamefont {Neupane},
  \citenamefont {Sanchez}, \citenamefont {Zheng}, \citenamefont {Jeng},
  \citenamefont {Bansil}, \citenamefont {Neupert}, \citenamefont {Lin},\ and\
  \citenamefont {Hasan}}]{Huang2016}%
  \BibitemOpen
  \bibfield  {author} {\bibinfo {author} {\bibfnamefont {S.-M.}\ \bibnamefont
  {Huang}}, \bibinfo {author} {\bibfnamefont {S.-Y.}\ \bibnamefont {Xu}},
  \bibinfo {author} {\bibfnamefont {I.}~\bibnamefont {Belopolski}}, \bibinfo
  {author} {\bibfnamefont {C.-C.}\ \bibnamefont {Lee}}, \bibinfo {author}
  {\bibfnamefont {G.}~\bibnamefont {Chang}}, \bibinfo {author} {\bibfnamefont
  {T.-R.}\ \bibnamefont {Chang}}, \bibinfo {author} {\bibfnamefont
  {B.}~\bibnamefont {Wang}}, \bibinfo {author} {\bibfnamefont {N.}~\bibnamefont
  {Alidoust}}, \bibinfo {author} {\bibfnamefont {G.}~\bibnamefont {Bian}},
  \bibinfo {author} {\bibfnamefont {M.}~\bibnamefont {Neupane}}, \bibinfo
  {author} {\bibfnamefont {D.}~\bibnamefont {Sanchez}}, \bibinfo {author}
  {\bibfnamefont {H.}~\bibnamefont {Zheng}}, \bibinfo {author} {\bibfnamefont
  {H.-T.}\ \bibnamefont {Jeng}}, \bibinfo {author} {\bibfnamefont
  {A.}~\bibnamefont {Bansil}}, \bibinfo {author} {\bibfnamefont
  {T.}~\bibnamefont {Neupert}}, \bibinfo {author} {\bibfnamefont
  {H.}~\bibnamefont {Lin}}, \ and\ \bibinfo {author} {\bibfnamefont {M.~Z.}\
  \bibnamefont {Hasan}},\ }\href {\doibase 10.1073/pnas.1514581113} {\bibfield
  {journal} {\bibinfo  {journal} {Proc. Natl. Acad. Sci}\ }\textbf {\bibinfo {volume} {113}},\
  \bibinfo {pages} {1180} (\bibinfo {year} {2016})}\BibitemShut {NoStop}%
\bibitem [{\citenamefont {Hirata}\ \emph {et~al.}(2016)\citenamefont {Hirata},
  \citenamefont {Ishikawa}, \citenamefont {Miyagawa}, \citenamefont {Tamura},
  \citenamefont {Berthier}, \citenamefont {Basko}, \citenamefont {Kobayashi},
  \citenamefont {Matsuno},\ and\ \citenamefont {Kanoda}}]{Hirata2016}%
  \BibitemOpen
  \bibfield  {author} {\bibinfo {author} {\bibfnamefont {M.}~\bibnamefont
  {Hirata}}, \bibinfo {author} {\bibfnamefont {K.}~\bibnamefont {Ishikawa}},
  \bibinfo {author} {\bibfnamefont {K.}~\bibnamefont {Miyagawa}}, \bibinfo
  {author} {\bibfnamefont {M.}~\bibnamefont {Tamura}}, \bibinfo {author}
  {\bibfnamefont {C.}~\bibnamefont {Berthier}}, \bibinfo {author}
  {\bibfnamefont {D.}~\bibnamefont {Basko}}, \bibinfo {author} {\bibfnamefont
  {A.}~\bibnamefont {Kobayashi}}, \bibinfo {author} {\bibfnamefont
  {G.}~\bibnamefont {Matsuno}}, \ and\ \bibinfo {author} {\bibfnamefont
  {K.}~\bibnamefont {Kanoda}},\ }\href {http://dx.doi.org/10.1038/ncomms12666}
  {\bibfield  {journal} {\bibinfo  {journal} {Nat. Commun.}\ }\textbf
  {\bibinfo {volume} {7}},\ \bibinfo {pages} {12666} (\bibinfo {year}
  {2016})}\BibitemShut {NoStop}%
\bibitem [{\citenamefont {Foster}\ and\ \citenamefont
  {Aleiner}(2008)}]{PhysRevB.77.195413}%
  \BibitemOpen
  \bibfield  {author} {\bibinfo {author} {\bibfnamefont {M.~S.}\ \bibnamefont
  {Foster}}\ and\ \bibinfo {author} {\bibfnamefont {I.~L.}\ \bibnamefont
  {Aleiner}},\ }\href {\doibase 10.1103/PhysRevB.77.195413} {\bibfield
  {journal} {\bibinfo  {journal} {Phys. Rev. B}\ }\textbf {\bibinfo {volume}
  {77}},\ \bibinfo {pages} {195413} (\bibinfo {year} {2008})}\BibitemShut
  {NoStop}%
\bibitem [{\citenamefont {Son}(2007)}]{PhysRevB.75.235423}%
  \BibitemOpen
  \bibfield  {author} {\bibinfo {author} {\bibfnamefont {D.~T.}\ \bibnamefont
  {Son}},\ }\href {\doibase 10.1103/PhysRevB.75.235423} {\bibfield  {journal}
  {\bibinfo  {journal} {Phys. Rev. B}\ }\textbf {\bibinfo {volume} {75}},\
  \bibinfo {pages} {235423} (\bibinfo {year} {2007})}\BibitemShut {NoStop}%
\bibitem [{\citenamefont {Landau}\ and\ \citenamefont
  {Lifshitz}(1998)}]{SWB-074898019}%
  \BibitemOpen
  \bibfield  {author} {\bibinfo {author} {\bibfnamefont {L.~D.}\ \bibnamefont
  {Landau}}\ and\ \bibinfo {author} {\bibfnamefont {E.~M.}\ \bibnamefont
  {Lifshitz}},\ }\href@noop {} {\emph {\bibinfo {title} {Fluid Mechanics}}},\
  \bibinfo {edition} {2nd}\ ed.\ (\bibinfo  {publisher}
  {Butterworth-Heinemann},\ \bibinfo {address} {Oxford},\ \bibinfo {year}
  {1998})\BibitemShut {NoStop}%
\bibitem [{\citenamefont {Bradlyn}\ \emph {et~al.}(2012)\citenamefont
  {Bradlyn}, \citenamefont {Goldstein},\ and\ \citenamefont
  {Read}}]{PhysRevB.86.245309}%
  \BibitemOpen
  \bibfield  {author} {\bibinfo {author} {\bibfnamefont {B.}~\bibnamefont
  {Bradlyn}}, \bibinfo {author} {\bibfnamefont {M.}~\bibnamefont {Goldstein}},
  \ and\ \bibinfo {author} {\bibfnamefont {N.}~\bibnamefont {Read}},\ }\href
  {\doibase 10.1103/PhysRevB.86.245309} {\bibfield  {journal} {\bibinfo
  {journal} {Phys. Rev. B}\ }\textbf {\bibinfo {volume} {86}},\ \bibinfo
  {pages} {245309} (\bibinfo {year} {2012})}\BibitemShut {NoStop}%
\bibitem [{\citenamefont {Principi}\ \emph {et~al.}(2016)\citenamefont
  {Principi}, \citenamefont {Vignale}, \citenamefont {Carrega},\ and\
  \citenamefont {Polini}}]{Principi2016}%
  \BibitemOpen
  \bibfield  {author} {\bibinfo {author} {\bibfnamefont {A.}~\bibnamefont
  {Principi}}, \bibinfo {author} {\bibfnamefont {G.}~\bibnamefont {Vignale}},
  \bibinfo {author} {\bibfnamefont {M.}~\bibnamefont {Carrega}}, \ and\
  \bibinfo {author} {\bibfnamefont {M.}~\bibnamefont {Polini}},\ }\href
  {\doibase 10.1103/PhysRevB.93.125410} {\bibfield  {journal} {\bibinfo
  {journal} {Phys. Rev. B}\ }\textbf {\bibinfo {volume} {93}},\ \bibinfo
  {pages} {125410} (\bibinfo {year} {2016})}\BibitemShut {NoStop}%
\bibitem [{\citenamefont {Sheehy}\ and\ \citenamefont
  {Schmalian}(2007)}]{Sheehy2007}%
  \BibitemOpen
  \bibfield  {author} {\bibinfo {author} {\bibfnamefont {D.~E.}\ \bibnamefont
  {Sheehy}}\ and\ \bibinfo {author} {\bibfnamefont {J.}~\bibnamefont
  {Schmalian}},\ }\href {\doibase 10.1103/PhysRevLett.99.226803} {\bibfield
  {journal} {\bibinfo  {journal} {Phys. Rev. Lett.}\ }\textbf {\bibinfo
  {volume} {99}},\ \bibinfo {pages} {226803} (\bibinfo {year}
  {2007})}\BibitemShut {NoStop}%
\bibitem [{\citenamefont {Hornreich}\ \emph {et~al.}(1975)\citenamefont
  {Hornreich}, \citenamefont {Luban},\ and\ \citenamefont
  {Shtrikman}}]{Hornreich1975}%
  \BibitemOpen
  \bibfield  {author} {\bibinfo {author} {\bibfnamefont {R.~M.}\ \bibnamefont
  {Hornreich}}, \bibinfo {author} {\bibfnamefont {M.}~\bibnamefont {Luban}}, \
  and\ \bibinfo {author} {\bibfnamefont {S.}~\bibnamefont {Shtrikman}},\ }\href
  {\doibase 10.1103/PhysRevLett.35.1678} {\bibfield  {journal} {\bibinfo
  {journal} {Phys. Rev. Lett.}\ }\textbf {\bibinfo {volume} {35}},\ \bibinfo
  {pages} {1678} (\bibinfo {year} {1975})}\BibitemShut {NoStop}%
\bibitem [{sup()}]{supplemental}%
  \BibitemOpen
  \bibinfo {note} {See Supplemetal Material at 
  \href {http://link.aps.org/supplemental/10.1103/PhysRevLett.120.196801} {http://link.aps.org/
  supplemental/10.1103/PhysRevLett.120.196801} for details
  of the calculations, which includes Refs.~[26,35,36].}\BibitemShut {Stop}%
\bibitem [{\citenamefont {Sch\"utt}\ \emph {et~al.}(2011)\citenamefont
  {Sch\"utt}, \citenamefont {Ostrovsky}, \citenamefont {Gornyi},\ and\
  \citenamefont {Mirlin}}]{RPA}%
  \BibitemOpen
  \bibfield  {author} {\bibinfo {author} {\bibfnamefont {M.}~\bibnamefont
  {Sch\"utt}}, \bibinfo {author} {\bibfnamefont {P.~M.}\ \bibnamefont
  {Ostrovsky}}, \bibinfo {author} {\bibfnamefont {I.~V.}\ \bibnamefont
  {Gornyi}}, \ and\ \bibinfo {author} {\bibfnamefont {A.~D.}\ \bibnamefont
  {Mirlin}},\ }\href {\doibase 10.1103/PhysRevB.83.155441} {\bibfield
  {journal} {\bibinfo  {journal} {Phys. Rev. B}\ }\textbf {\bibinfo {volume}
  {83}},\ \bibinfo {pages} {155441} (\bibinfo {year} {2011})}\BibitemShut
  {NoStop}%
\bibitem [{\citenamefont {M\"uller}\ \emph {et~al.}(2009)\citenamefont
  {M\"uller}, \citenamefont {Schmalian},\ and\ \citenamefont
  {Fritz}}]{PhysRevLett.103.025301}%
  \BibitemOpen
  \bibfield  {author} {\bibinfo {author} {\bibfnamefont {M.}~\bibnamefont
  {M\"uller}}, \bibinfo {author} {\bibfnamefont {J.}~\bibnamefont {Schmalian}},
  \ and\ \bibinfo {author} {\bibfnamefont {L.}~\bibnamefont {Fritz}},\ }\href
  {\doibase 10.1103/PhysRevLett.103.025301} {\bibfield  {journal} {\bibinfo
  {journal} {Phys. Rev. Lett.}\ }\textbf {\bibinfo {volume} {103}},\ \bibinfo
  {pages} {025301} (\bibinfo {year} {2009})}\BibitemShut {NoStop}%
\bibitem [{\citenamefont {Briskot}\ \emph {et~al.}(2015)\citenamefont
  {Briskot}, \citenamefont {Sch\"utt}, \citenamefont {Gornyi}, \citenamefont
  {Titov}, \citenamefont {Narozhny},\ and\ \citenamefont {Mirlin}}]{us2}%
  \BibitemOpen
  \bibfield  {author} {\bibinfo {author} {\bibfnamefont {U.}~\bibnamefont
  {Briskot}}, \bibinfo {author} {\bibfnamefont {M.}~\bibnamefont {Sch\"utt}},
  \bibinfo {author} {\bibfnamefont {I.~V.}\ \bibnamefont {Gornyi}}, \bibinfo
  {author} {\bibfnamefont {M.}~\bibnamefont {Titov}}, \bibinfo {author}
  {\bibfnamefont {B.~N.}\ \bibnamefont {Narozhny}}, \ and\ \bibinfo {author}
  {\bibfnamefont {A.~D.}\ \bibnamefont {Mirlin}},\ }\href {\doibase
  10.1103/PhysRevB.92.115426} {\bibfield  {journal} {\bibinfo  {journal} {Phys.
  Rev. B}\ }\textbf {\bibinfo {volume} {92}},\ \bibinfo {pages} {115426}
  (\bibinfo {year} {2015})}\BibitemShut {NoStop}%
\bibitem [{\citenamefont {M\"uller}\ \emph {et~al.}(2008)\citenamefont
  {M\"uller}, \citenamefont {Fritz},\ and\ \citenamefont
  {Sachdev}}]{PhysRevB.78.115406}%
  \BibitemOpen
  \bibfield  {author} {\bibinfo {author} {\bibfnamefont {M.}~\bibnamefont
  {M\"uller}}, \bibinfo {author} {\bibfnamefont {L.}~\bibnamefont {Fritz}}, \
  and\ \bibinfo {author} {\bibfnamefont {S.}~\bibnamefont {Sachdev}},\ }\href
  {\doibase 10.1103/PhysRevB.78.115406} {\bibfield  {journal} {\bibinfo
  {journal} {Phys. Rev. B}\ }\textbf {\bibinfo {volume} {78}},\ \bibinfo
  {pages} {115406} (\bibinfo {year} {2008})}\BibitemShut {NoStop}%
\bibitem [{\citenamefont {Foster}\ and\ \citenamefont
  {Aleiner}(2009)}]{Foster2009}%
  \BibitemOpen
  \bibfield  {author} {\bibinfo {author} {\bibfnamefont {M.~S.}\ \bibnamefont
  {Foster}}\ and\ \bibinfo {author} {\bibfnamefont {I.~L.}\ \bibnamefont
  {Aleiner}},\ }\href {\doibase 10.1103/PhysRevB.79.085415} {\bibfield
  {journal} {\bibinfo  {journal} {Phys. Rev. B}\ }\textbf {\bibinfo {volume}
  {79}},\ \bibinfo {pages} {085415} (\bibinfo {year} {2009})}\BibitemShut
  {NoStop}%
\bibitem [{\citenamefont {Narozhny}\ \emph {et~al.}(2015)\citenamefont
  {Narozhny}, \citenamefont {Gornyi}, \citenamefont {Titov}, \citenamefont
  {Sch\"utt},\ and\ \citenamefont {Mirlin}}]{PhysRevB.91.035414}%
  \BibitemOpen
  \bibfield  {author} {\bibinfo {author} {\bibfnamefont {B.~N.}\ \bibnamefont
  {Narozhny}}, \bibinfo {author} {\bibfnamefont {I.~V.}\ \bibnamefont
  {Gornyi}}, \bibinfo {author} {\bibfnamefont {M.}~\bibnamefont {Titov}},
  \bibinfo {author} {\bibfnamefont {M.}~\bibnamefont {Sch\"utt}}, \ and\
  \bibinfo {author} {\bibfnamefont {A.~D.}\ \bibnamefont {Mirlin}},\ }\href
  {\doibase 10.1103/PhysRevB.91.035414} {\bibfield  {journal} {\bibinfo
  {journal} {Phys. Rev. B}\ }\textbf {\bibinfo {volume} {91}},\ \bibinfo
  {pages} {035414} (\bibinfo {year} {2015})}\BibitemShut {NoStop}%
\end{thebibliography}

\begin{thebibliography}{4}%
\makeatletter
\providecommand \@ifxundefined [1]{%
 \@ifx{#1\undefined}
}%
\providecommand \@ifnum [1]{%
 \ifnum #1\expandafter \@firstoftwo
 \else \expandafter \@secondoftwo
 \fi
}%
\providecommand \@ifx [1]{%
 \ifx #1\expandafter \@firstoftwo
 \else \expandafter \@secondoftwo
 \fi
}%
\providecommand \natexlab [1]{#1}%
\providecommand \enquote  [1]{``#1''}%
\providecommand \bibnamefont  [1]{#1}%
\providecommand \bibfnamefont [1]{#1}%
\providecommand \citenamefont [1]{#1}%
\providecommand \href@noop [0]{\@secondoftwo}%
\providecommand \href [0]{\begingroup \@sanitize@url \@href}%
\providecommand \@href[1]{\@@startlink{#1}\@@href}%
\providecommand \@@href[1]{\endgroup#1\@@endlink}%
\providecommand \@sanitize@url [0]{\catcode `\\12\catcode `\$12\catcode
  `\&12\catcode `\#12\catcode `\^12\catcode `\_12\catcode `\%12\relax}%
\providecommand \@@startlink[1]{}%
\providecommand \@@endlink[0]{}%
\providecommand \url  [0]{\begingroup\@sanitize@url \@url }%
\providecommand \@url [1]{\endgroup\@href {#1}{\urlprefix }}%
\providecommand \urlprefix  [0]{URL }%
\providecommand \Eprint [0]{\href }%
\providecommand \doibase [0]{http://dx.doi.org/}%
\providecommand \selectlanguage [0]{\@gobble}%
\providecommand \bibinfo  [0]{\@secondoftwo}%
\providecommand \bibfield  [0]{\@secondoftwo}%
\providecommand \translation [1]{[#1]}%
\providecommand \BibitemOpen [0]{}%
\providecommand \bibitemStop [0]{}%
\providecommand \bibitemNoStop [0]{.\EOS\space}%
\providecommand \EOS [0]{\spacefactor3000\relax}%
\providecommand \BibitemShut  [1]{\csname bibitem#1\endcsname}%
\let\auto@bib@innerbib\@empty
\bibitem [{\citenamefont {Landau}\ and\ \citenamefont
  {Lifshitz}(1998)}]{SWB-074898019}%
  \BibitemOpen
  \bibfield  {author} {\bibinfo {author} {\bibfnamefont {L.~D.}\ \bibnamefont
  {Landau}}\ and\ \bibinfo {author} {\bibfnamefont {E.~M.}\ \bibnamefont
  {Lifshitz}},\ }\href@noop {} {\emph {\bibinfo {title} {Fluid mechanics}}},\
  \bibinfo {edition} {2nd}\ ed.\ (\bibinfo  {publisher}
  {Butterworth-Heinemann},\ \bibinfo {address} {Oxford},\ \bibinfo {year}
  {1998})\BibitemShut {NoStop}%
\bibitem [{\citenamefont {Isobe}\ \emph {et~al.}(2016)\citenamefont {Isobe},
  \citenamefont {Yang}, \citenamefont {Chubukov}, \citenamefont {Schmalian},\
  and\ \citenamefont {Nagaosa}}]{PhysRevLett.116.076803}%
  \BibitemOpen
  \bibfield  {author} {\bibinfo {author} {\bibfnamefont {H.}~\bibnamefont
  {Isobe}}, \bibinfo {author} {\bibfnamefont {B.-J.}\ \bibnamefont {Yang}},
  \bibinfo {author} {\bibfnamefont {A.}~\bibnamefont {Chubukov}}, \bibinfo
  {author} {\bibfnamefont {J.}~\bibnamefont {Schmalian}}, \ and\ \bibinfo
  {author} {\bibfnamefont {N.}~\bibnamefont {Nagaosa}},\ }\href {\doibase
  10.1103/PhysRevLett.116.076803} {\bibfield  {journal} {\bibinfo  {journal}
  {Phys. Rev. Lett.}\ }\textbf {\bibinfo {volume} {116}},\ \bibinfo {pages}
  {076803} (\bibinfo {year} {2016})}\BibitemShut {NoStop}%
\bibitem [{\citenamefont {Bradlyn}\ \emph {et~al.}(2012)\citenamefont
  {Bradlyn}, \citenamefont {Goldstein},\ and\ \citenamefont
  {Read}}]{PhysRevB.86.245309}%
  \BibitemOpen
  \bibfield  {author} {\bibinfo {author} {\bibfnamefont {B.}~\bibnamefont
  {Bradlyn}}, \bibinfo {author} {\bibfnamefont {M.}~\bibnamefont {Goldstein}},
  \ and\ \bibinfo {author} {\bibfnamefont {N.}~\bibnamefont {Read}},\ }\href
  {\doibase 10.1103/PhysRevB.86.245309} {\bibfield  {journal} {\bibinfo
  {journal} {Phys. Rev. B}\ }\textbf {\bibinfo {volume} {86}},\ \bibinfo
  {pages} {245309} (\bibinfo {year} {2012})}\BibitemShut {NoStop}%
\end{thebibliography}

%

    \end{document}